\documentclass[aps,prb,twocolumn,superscriptaddress,10pt,article,nofootinbib,showpacs,longbibliography]{revtex4-1}
\usepackage{amsfonts}
\usepackage{amsmath}
\usepackage{amsthm}
\usepackage{braket}
\usepackage{graphicx}
\usepackage{hyperref}
\hypersetup{
    colorlinks=true,
    linkcolor=blue,
    filecolor=cyan,      
    urlcolor=magenta,
    citecolor=purple,
}
\usepackage[dvipsnames]{xcolor}
\usepackage{amssymb}
\usepackage{dsfont}
\usepackage{verbatim}
\usepackage{amsmath,amscd}
\usepackage{multirow}
\usepackage[normalem]{ulem}
\usepackage{bm}
\def \k{{\mathbf k}}
\def \q{{\mathbf q}}

\def \A{{\mathbf A}}

\def \d{{\bm d}}
\def \n{{\mathbf n}}
\def \j{{\mathbf j}}
\def \r{{\mathbf r}}

\def \v{{\mathbf v}}
\def \A{{\mathbf A}}
\def \B{{\mathbf B}}
\def \E{{\mathbf E}}

\def \N{{\mathcal N}}

\def \ua{\uparrow}
\def \da{\downarrow}
\def \beq{\begin{eqnarray}}
\def \eeq{\end{eqnarray}}
\def \nn{\nonumber \\}

\DeclareMathOperator{\Tr}{Tr}

\begin{document}

\title{Characterizing two-dimensional superconductivity via nanoscale noise magnetometry
with single-spin qubits}

\author{Pavel E. Dolgirev}
\thanks{P.E.D. and S.C. contributed equally to this work.}
\affiliation{Department of Physics, Harvard University, Cambridge, Massachusetts 02138, USA}

\author{Shubhayu Chatterjee}
\thanks{P.E.D. and S.C. contributed equally to this work.}
\affiliation{Department of Physics, University of California, Berkeley, CA 94720, USA}

\author{Ilya Esterlis}
\affiliation{Department of Physics, Harvard University, Cambridge, Massachusetts 02138, USA}
\author{Alexander~A.~Zibrov}
\affiliation{Department of Physics, Harvard University, Cambridge, Massachusetts 02138, USA}
\author{Mikhail~D.~Lukin}
\affiliation{Department of Physics, Harvard University, Cambridge, Massachusetts 02138, USA}
\author{Norman~Y.~Yao}
\affiliation{Department of Physics, University of California, Berkeley, CA 94720, USA}
\affiliation{Materials Sciences Division, Lawrence Berkeley National Laboratory, Berkeley CA 94720, USA}
\author{Eugene Demler}
\affiliation{Department of Physics, Harvard University, Cambridge, Massachusetts 02138, USA}
\affiliation{Institute for Theoretical Physics, ETH Zurich, 8093 Zurich, Switzerland}

\begin{abstract}
We propose nanoscale magnetometry via isolated single-spin qubits as a probe of superconductivity in two-dimensional materials. 
We characterize the magnetic field noise at the qubit location, arising from current and spin fluctuations in the sample and leading to measurable polarization decay of the qubit. 
We show that the noise due to transverse current fluctuations studied as a function of temperature and sample-probe distance can be used to extract useful information about the transition to a superconducting phase and the pairing symmetry of the superconductor. 
Surprisingly, at low temperatures, the dominant contribution to the magnetic noise arises from longitudinal current fluctuations and can be used to probe collective modes such as monolayer plasmons and bilayer Josephson plasmons.
We also characterize the noise due to spin fluctuations, which allows probing the spin structure of the pairing wave function. Our results provide a non-invasive route to probe the rich physics of two-dimensional superconductors.
\end{abstract}

\maketitle

\section{Introduction}
\label{sec:intro}
Recent years have witnessed a surge of activity on two-dimensional (2D) superconductors on both experimental and theoretical fronts. 
On the experimental side, robust superconductivity has been observed in transport measurements in several 2D materials, including Van der Waals heterostructures such as magic angle graphene and transition metal dichalcogenides (TMDs)~\cite{cao2018,lu2019,yankowitz2019,park2021,hao2021,shi2015}. 
On the theoretical side, analytical and numerical studies have predicted both new material candidates and new physical mechanisms for 2D superconductivity~\cite{CIZ2020,Song2021,Scheurer,CBZ2020,Gonzalez2018,Chichinadze2020,KC2020,Christos}. 
However, for some exciting prospective 2D materials such as TMDs, it is experimentally challenging to make electric contacts that are necessary to carry out transport measurements to detect a superconducting phase transition~\cite{Allain2015}. For magic-angle twisted bilayer or trilayer graphene, resistivity measurements do show a superconducting transition at low temperatures. 
However, the nature of the superconductivity there, as well as the symmetry of the gap function, remains unknown, as it is not unambiguously accessible with conventional probes. 
Therefore, it is highly desirable to devise complementary experimental probes that can efficiently detect and characterize 2D superconductivity. 

Quantum sensing has established itself as a rapidly growing area of research, with tremendous technological prospects \cite{Degen_RMP}. 
In particular, isolated impurity qubits, such as nitrogen-vacancy (NV) or silicon-vacancy (SiV) centers in diamond, have enabled measurements of local magnetic fields with high precision and accuracy \cite{hong2013nanoscale,nvreview,nvsinglespin,Casola,Stano2013}.
While such qubits have broad applications ranging from quantum computation to biological imaging~\cite{acosta_hemmer_2013}, very recently, they have also proven useful in understanding the behavior of condensed matter systems. Both static sensing of local magnetic fields and dynamic detection of magnetic noise have been used to study interesting physics, including topological magnetic textures such as skyrmions \cite{Dovzhenko2018}, non-local transport in metals \cite{Kolkowitz}, pressure-driven phase transitions \cite{Satcher,lesik2019}, and scattering of magnons in magnetic thin films \cite{zhou2020magnon}, to name a few. 
Bolstered by this, there have been theoretical proposals to use magnetic noise sensors to probe a variety of phenomena, such as symmetry-protected one-dimensional edge modes \cite{Joaquin18}, hydrodynamic sound modes in magnon fluids \cite{Joaquin_magnon_18}, dynamic phase transitions via magnon condensation \cite{FT2018}, and exotic long-range entangled states such as quantum spin liquids \cite{CRD18}.
The qubit sensors offer several advantages compared to traditional condensed matter probes. 
Their optical initialization and read-out capabilities, high degree of tunability with both frequency and momentum resolution, and minimally invasive nature make them ideal for characterizing the physics of correlated electronic systems.

In this work, we propose nanoscale noise magnetometry by impurity qubits as a probe of 2D superconductivity.
When an isolated qubit is placed in proximity to a 2D superconductor, it couples to the noisy magnetic field generated by both current and spin fluctuations in the sample. If the qubit is initialized in a polarized state, the polarization will decay due to magnetic noise. It is convenient to distinguish transverse current fluctuations, where the current density $ \j^{\rm T}(\q)$ is perpendicular to the in-plane momentum $\q$, from longitudinal current fluctuations, with $\j^{\rm L}(\q) \parallel \q$. The latter are accompanied by charge density fluctuations, as follows from the continuity equation, and, thus, are suppressed by strong Coulomb forces. For this reason, the longitudinal sector can be safely neglected in metals~\cite{Agarwal2017}. In superconductors, this is no longer true at low temperatures, where the presence of superconductivity suppresses transverse current fluctuations, thus providing a gateway to probe the longitudinal ones. In Sec.~\ref{sec:Rel}, we discuss how one gains independent access to both transverse and longitudinal sectors by varying the direction of the initial qubit polarization.

In Sec.~\ref{sec:NT}, we demonstrate that within the two-fluid model of superconductors, the transverse magnetic noise is essentially determined by the transverse conductivity $\sigma_n^{\rm T}(\q, \Omega)$ of the normal fluid. The frequency $\Omega$ is set by the probe-splitting and can be tuned by external fields, while the in-plane momentum $q \sim 1/z_0$ is set by the inverse sample-probe distance. Typically, the qubit energy splitting $\Omega$ is in the gigahertz range, being the smallest energy scale in the system, allowing approximating $\Omega \approx 0$. At the same time, what makes qubit sensors distinct compared to conventional probes is the tunability of the sample-probe distance $z_0$, giving access to various transport regimes, as encoded in $\sigma_n^{\rm T}(q\sim 1/z_0, \Omega \to 0)$. These possible transport regimes are determined by the rich interplay of four length scales in superconductors: (i) the quasiparticle mean-free path $\ell_{\rm MF}$, (ii) the qubit-sample distance $z_0$, (iii) the quasiparticle thermal wavelength $\lambda_T$, and (iv) the superconducting coherence length $\xi_T$. In Sec.~\ref{sec:NT}, to investigate these transport regimes as well as crossovers between them that occur upon tuning experimental knobs (for instance, the temperature $T$), we compute the transverse normal conductivity within the mean-field BCS theory using the Kubo formula. This one-loop calculation neglects the long-range Coulomb interaction, which, however, is not expected to affect $\sigma_n^{\rm T}$. We examine both singlet and triplet superconductors, with different symmetries of the superconducting order parameter, in both clean and disordered limits. The main result of Sec.~\ref{sec:NT} is the demonstration that qubit sensors can be used to detect the superconducting phase transition and to uncover the nature of the pairing function.

\begin{figure}
    \centering
    \includegraphics[width=0.4\textwidth]{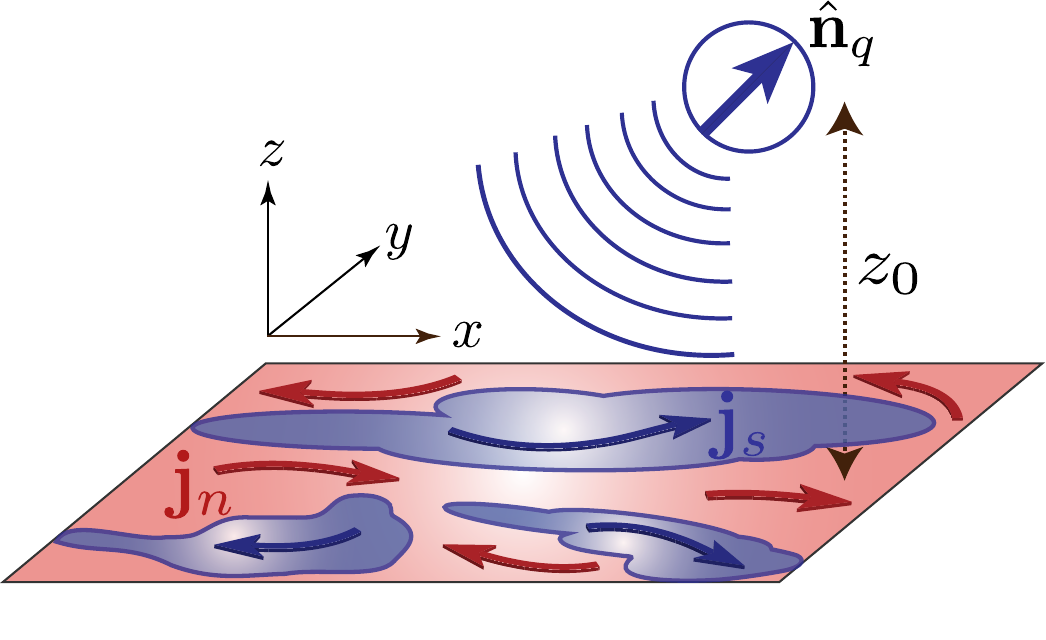}
    \caption{Schematic of the experimental set-up, showing an isolated impurity qubit placed at a distance $z_0$ from the two-dimensional superconducting sample in the $x$-$y$ plane. A fluctuating magnetic field due to both supercurrent $\j_s$ and normal current $\j_n$ in the sample results in relaxation of the qubit polarization.}
    \label{fig:Schematic}
\end{figure}

Deep in the superconducting phase, quasiparticle excitations become thermally suppressed due to the superconducting gap, leading to the suppression of the transverse noise. At such low temperatures, one no longer can neglect longitudinal current fluctuations. In Sec.~\ref{sec:NL}, we investigate the longitudinal noise and show that it allows us to probe longitudinal collective modes, such as gapless plasmons in monolayers and gapped Josephson plasmons in bilayers.

In addition to current fluctuations, spin fluctuations can also contribute to the magnetic noise.
The suppression of the transverse current fluctuations at low temperatures requires us to address the question of spin noise carefully, which we do in Sec.~\ref{sec:SpinN}. We find that in contrast to metals, spin noise is not parametrically suppressed as a function of the sample-probe distance, but its magnitude is still quite small. It may become comparable to the current noise in systems with flat bands or with bands having large Berry curvature relevant to some moir\'{e} materials, in which case the anisotropy of noise can be used to determine the nature of triplet pairing.

We conclude and give a brief outlook for future work in Sec.~\ref{sec:conc}. Technical calculations are relegated to appendixes. Some highlights of this work can be found in the shorter paper, Ref.~\onlinecite{Shubhayu1}.

\section{Relaxation rate of qubit}
\label{sec:Rel}

We begin by characterizing the depolarization of the qubit in the presence of a nearby superconducting sample. For concreteness, we consider a single isolated qubit at $\r_0 = (0,0,z_0)$, i.e, at a distance $z_0$ above the two-dimensional homogeneous sample in the $xy$ plane. The qubit Hamiltonian is given by a splitting $\Omega$ along a quantization axis $\hat{\n}_q$ and a coupling to the local magnetic field $\B(\r_0,t)$ at the qubit location:
\beq
H_q =  \frac{\Omega}{2}(\hat{\n}_q \cdot \bm{\sigma}) + g \mu_B \B(\r_0, t) \cdot \bm{\sigma}.
\eeq
The magnetic field $\B(\r_0,t)$ comes from charge and spin fluctuations in the sample.
Once the qubit is initialized in a polarized state, the qubit polarization will decay due to the coupling to this noisy field. 
This can be characterized by the magnetic noise tensor $\N_{ab}(\Omega)$:
\beq
\N_{ab}(\Omega) = \frac{1}{2} \int_{-\infty}^{\infty} dt \ e^{i \Omega t} \langle \{ B_a(\r_0,t) , B_b(\r_0,0) \} \rangle,
\eeq
where $\langle\,\dots\,\rangle$ denotes equilibrium ensemble average at temperature $T$. By a standard application of Fermi's golden rule (see Refs.~\onlinecite{Joaquin18,CRD18} for details), the relaxation rate of the qubit polarization can be related to the noise tensor as:
\beq
\frac{1}{T_{1}} = (g \mu_B)^2 \N_{+-}(\Omega),
\label{eq:Tinv}
\eeq
where $\B_{\pm} = B_{x^\prime} \pm i B_{y^\prime}$, and $(\hat{x}^\prime,\hat{y}^\prime, \hat{\n}_q)$ form a mutually orthogonal triad (the qubit quantization axis $\hat{\n}_q$ need not to coincide with the $z$-axis, as depicted in Fig.~\ref{fig:Schematic}).

Useful constraints on the magnetic noise tensor can be derived from symmetry considerations. Assuming the rotational symmetry about the $z$ axis, we get $\N_{xx} = \N_{yy}$ and $\N_{xy} = - \N_{yx}$. 
On additional imposition of reflection symmetry in the $xz$ ($yz$) plane, we find that $\N_{xz (yz)} = 0$.
Therefore, the noise tensor at a given frequency $\Omega$ is completely characterized by two independent numbers, namely: the transverse noise $\N_{\rm T} = \N_{xx} + \N_{yy} = 2 \N_{xx}$ and the longitudinal noise $\N_{\rm L} = \N_{zz}$. 
From Eq.~\eqref{eq:Tinv}, we note that the orientation of $\hat{\n}_q$ tells us what kind of noise the qubit is sensitive to.
Specifically, setting $\hat{\n}_q = \hat{z}$ makes the qubit sensitive to $\N_{\rm T}$, while setting $\hat{\n}_q = \hat{x}$ results in a relaxation time governed by $\N_{\rm L} + \N_{\rm T}/2$.
Therefore, both transverse and longitudinal noise can be extracted independently by appropriate alignment of the qubit quantization axis.

\begin{figure}
    \centering
    \includegraphics[width=0.48\textwidth]{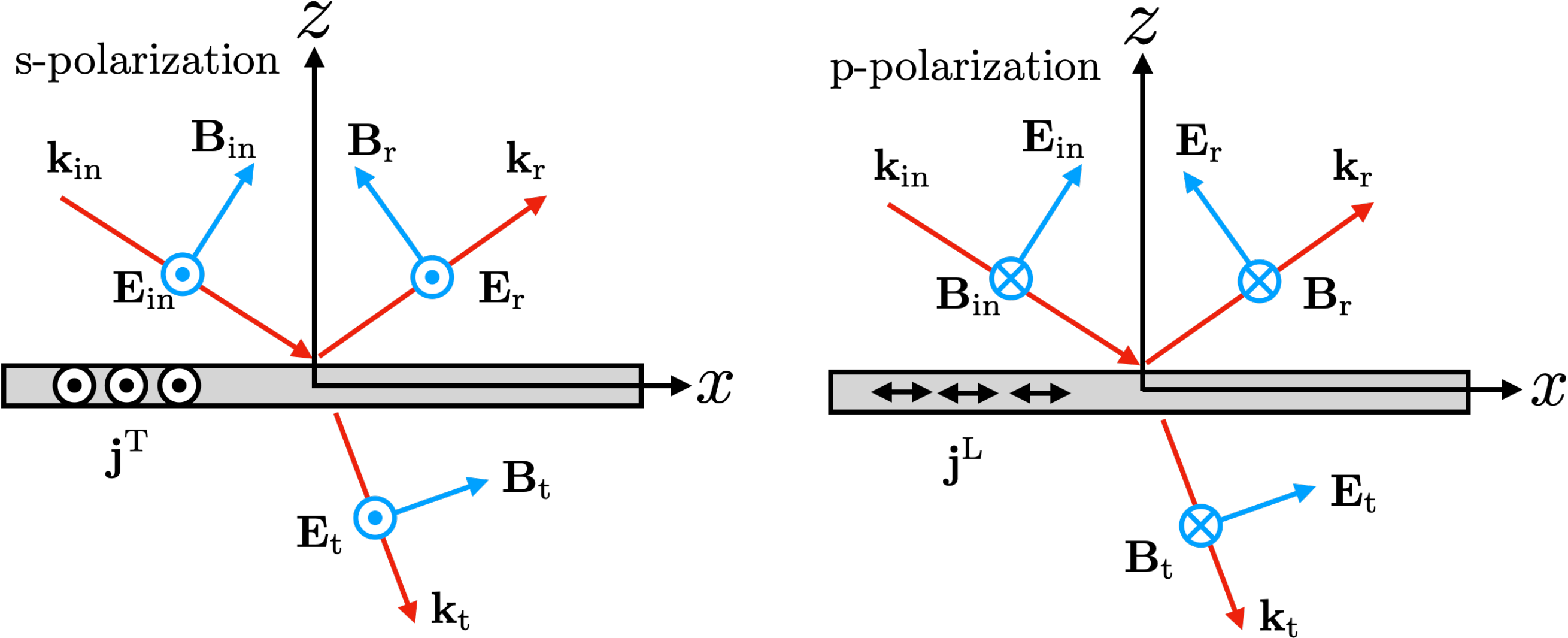}
    \caption{Definition of s- and p-polarized waves. Subscripts ``in",``r", and ``t", denote ``incoming", "reflected", and ``transmitted" waves, respectively. The s-polarization generates transverse currents, $\j^{\rm T}$, while p-polarization generates longitudinal currents, $\j^{\rm L}$.}
    \label{fig:Polarizations}
\end{figure}

What remains is to relate the noise tensor $\N_{ab}$ to correlations within the sample.
As discussed in Ref.~\onlinecite{Agarwal2017}, this can be done by solving Maxwell's equations, which relate the magnetic field at $\r_{0}$ to fluctuating sources in the superconductor. 
Neglecting retardation effects (since the speed of light $c$ is much larger than typical velocity scales, such as Fermi velocity $v_F$, in condensed matter systems), most of the noise comes from evanescent electromagnetic (EM) modes.
In particular, the transverse noise $\N_{\rm T}$ is given by:
\beq
\N_{\rm T}(\Omega) = \frac{\mu_0 k_B T}{16 \pi \Omega z_0^3} \int_0^\infty dx \, x^2 e^{-x} \text{Im} \left[ r_s\left( \frac{x}{2z_0},\Omega \right) \right], ~~
\label{eq:NT}
\eeq
where $r_s(\q,\Omega)$ is the reflection coefficient for s-polarized EM waves which couple to transverse currents
($\q \cdot \j^{\rm T}(\q) = 0$), see Fig.~\ref{fig:Polarizations}. Here $\q$ is the in-plane momentum and the out-of-plane momentum is substituted with $q_z \approx iq$ (so that we consider only evanescent waves).
Therefore, if we decompose the conductivity tensor into transverse and longitudinal components as $\sigma_{ab}(\q, \Omega) = \sigma^{\rm T}(\q,\Omega) \left(\delta_{ab} - \frac{q_a q_b}{q^2}  \right) + \sigma^{\rm L}(\q,\Omega) \frac{q_a q_b}{q^2} $, the reflection coefficient $r_s$ can be written in terms of the transverse conductivity $\sigma^{\rm T}(\q,\Omega)$ as follows \cite{Agarwal2017}: 
\beq
r_s(\q,\Omega) = -\left(1 + \frac{2i q}{\mu_0 \Omega \, \sigma^{\rm T}(\q,\Omega)}\right)^{-1}.
\label{eq:rs}
\eeq
In an analogous manner, the longitudinal noise $\N_{\rm L}$ is related to the reflection coefficient $r_p(\q,\Omega)$ of p-polarized electromagnetic waves (see Fig.~\ref{fig:Polarizations}):
\beq
\N_{\rm L}(\Omega) = \frac{\mu_0 k_B T }{8 \pi \Omega z_0^3} \left( \frac{\Omega z_0}{c} \right)^2 \int_0^\infty dx \, e^{-x} \text{Im} \left[ r_p\left( \frac{x}{2z_0},\Omega \right) \right], \nn
\label{eq:NL}
\eeq
where $r_p$ is related to $\sigma^{\rm L}(\q,\Omega)$ according to \cite{Agarwal2017}:
\beq
r_p(\q, \Omega)= \left(  1 + \frac{2 \epsilon \epsilon_0 \Omega}{i q \sigma^{\rm L}(\q,\Omega)} \right)^{-1}.
\label{eq:rp}
\eeq
The additional suppression factor of $(\Omega z_0/c)^2$ for longitudinal noise is due to the fact that p-polarized waves couple to longitudinal currents, and hence charge density fluctuations, which are efficiently screened if the 2D sample is a good conductor.
For typical values of $\Omega \approx 10$ GHz and $z_0 \approx 100$ nm, $\Omega z_0/c \approx 10^{-5}$.
In metals, one therefore expects $\N_{\rm L}$ to be highly suppressed relative to $\N_{\rm T}$, so that it is safe to neglect its contribution to the qubit relaxation rate, as was done in Ref.~\onlinecite{Agarwal2017}.  
Contrary to this intuition, we will find that this is no longer the case at low temperatures in superconductors, when $\N_{\rm T}$ becomes suppressed due to the spectral gap, while $\N_{\rm L}$ can be resonantly enhanced by collective modes. 
The task at hand is now clear from Eqs.~\eqref{eq:NT}-\eqref{eq:rp}: we need to compute the non-local conductivity $\sigma^{\rm T/L}(\q,\Omega)$ for the 2D superconducting sample we are interested in. 
To do so, we need to model superconductivity to account for both quasiparticle and superflow contributions, as we discuss in Sec. \ref{sec:NT}.

Before switching to a detailed evaluation of the magnetic noise due to current fluctuations, we attempt to gain some intuitive understanding of how it scales with the qubit-probe distance $z_0$ \cite{Agarwal2017}.
The magnetic noise sensed by the qubit probe is proportional to $|\B(\r_0)|^2$. This local magnetic field is related to the current in the sample through the Biot-Savart kernel $K_j(z_0) \sim 1/z_0^2$.
The qubit probe is most sensitive to current fluctuations occurring at length-scales of $z_0$ (at higher momenta they are suppressed by the evanescent nature of EM waves carrying the signal, whereas at lower momenta they are suppressed due to low phase space in 2D). 
This approximately corresponds to \textit{seeing} an area ${\cal A}\sim z_0^2$ of the sample.
Thus, we can estimate the noise due to current fluctuations as (defining $\r_{0i} = \r_0 - \r_i, i = 1,2$, and using $|\r_{0i}| \approx z_0$):
\beq
\N_{\rm T/L} &  \approx & \mu_0^2 \int_{\cal{A}} d^2\r_1 \int_{\cal{A}} d^2\r_2 K_j(\r_{01}) K_j(\r_{02})\nn
&&
\qquad\qquad\qquad\qquad\qquad
\times\langle \{  j_{\rm T/L}(\r_1),  j_{\rm T/L}(\r_2) \} \rangle \nn
& \approx & \frac{\mu_0^2}{z_0^{4}} \int_{\cal{A}} d^2\mathbf{R} \int_{\cal{A}} d^2\r \langle\{ j_{\rm T/L}(\r),j_{\rm T/L}(0) \} \rangle \nn 
&=& \frac{\mu_0^2}{z_0^{2}} \int_{\cal{A}} d^2\r \langle \{ j_{\rm T/L}(\r),j_{\rm T/L}(0) \} \rangle.
\label{eq:SchemNTL}
\eeq
In the last step, we have used the translational invariance of the current correlations to separate the integration into center of mass and relative coordinates ($\mathbf{R} = (\r_1 + \r_2)/2$ and $\r = \r_1 - \r_2$, respectively), both of which are integrated over areas of linear dimensions $z_0$ of the sample.
In the simplest scenario, the correlation length is smaller than $z_0$, so that the $\r$ integral yields a finite value independent of $z_0$, implying that $\N\sim1/z_0^2$. 
More broadly, the noise scaling with distance, which is generically different from $\sim 1/z_0^2$, contains essential information about the current-current correlation function and, thus, the conductivity, as we will encounter in the subsequent sections. 
We remark that similar analysis as in Eq.~\eqref{eq:SchemNTL} will prove useful to understand the noise due to spin fluctuations in the sample, as we demonstrate in Sec.~\ref{sec:SpinN}.

\section{Noise in transverse sector}
\label{sec:NT}

We turn to discuss the transverse noise, which is determined by the transverse conductivity $\sigma^{\rm T}(\q,\Omega)$, via Eqs.~\eqref{eq:NT} and \eqref{eq:rs}. To evaluate $\sigma^{\rm T}(\q,\Omega)$, we employ the two-fluid model of superconductors~\cite{Tinkham}, which divides the total electron density into a superfluid density $n_s$ and a normal-fluid density $n_n$, as illustrated in Fig.~\ref{fig:Schematic}.
Accordingly, the total current density $\j$ is given by the sum of the normal-fluid contribution $\j_n = \sigma_n \E$ and the superfluid contribution $\j_s = \sigma_s \E$, so that the net conductivity is $\sigma^{\rm T}(\q,\Omega) = \sigma^{\rm T}_s(\q,\Omega) + \sigma_n^{\rm T}(\q,\Omega)$. The superfluid response is reactive, as follows from London's equation:
\begin{equation}
    \j_s^{\rm T} = - \Lambda \A^{\rm T} 
    \implies \sigma^{\rm T}_s(\q,\Omega) = -\frac{\Lambda}{i \Omega},
\end{equation}
where $\A^{\rm T}$ is the vector potential in the London gauge, satisfying $\q \cdot \A^{\rm T}(\q,\Omega) = 0$. Within the phenomenological Landau-Ginzburg theory of superconductivity, we identify: 
\begin{equation}
    \Lambda \propto n_{s} \propto |\Delta(T)|^2 \propto T_c - T, \text{ for } T < T_c. 
\label{eq:Lambda}
\end{equation}
Here $n_{s} = d|\psi|^2$ is the 2D superfluid density ($d$ being the sample thickness and $\psi$ being the superconducting order parameter), and $\Delta = g_{\rm BCS}|\psi|$ is the quasiparticle gap related to the order parameter through the effective attractive BCS electron-electron coupling $g_{\rm BCS}$.
The net transverse electrical conductivity is therefore given by:
\beq
\sigma^{\rm T}(\q,\Omega) = \sigma^{\rm T}_n(\q,\Omega) - \frac{\Lambda}{i \Omega}
\label{eq:SigmaT}.
\eeq
Before turning to the computation of the normal conductivity $\sigma^{\rm T}_n$, we first  discuss the effect of the superflow on the transverse noise. 
Plugging in Eq.~\eqref{eq:SigmaT} into Eqs.~\eqref{eq:NT} and~\eqref{eq:rs}, we obtain:
\beq
\mathcal{N}_{\rm T}(\Omega) & \approx & \frac{ \mu_0^2 k_B T}{16 \pi z^2_{0}} \int_0^\infty dx \, \frac{x^3 e^{-x} \, \text{Re}\left\{ \sigma_n^{\rm T}\Big(\frac{x}{2z_0},\Omega \Big) \right\} }{(\mu_0 z_0 \Omega \sigma^{\rm T}_n)^2 + (x + \mu_0 z_{0} \Lambda)^2} \nn
& \approx & \frac{ \mu_0^2 k_B T}{16 \pi z^2_{0}} \int_0^\infty dx \, x^3 e^{-x} \frac{\text{Re}\left\{ \sigma_n^{\rm T}\Big(\frac{x}{2z_0},\Omega \Big) \right\} }{(x + \mu_0 z_{0} \Lambda)^2}, \label{eqn:N_T_gen}
\eeq
where we approximated $\Omega \approx 0$ in the denominator in the first line, since the probe-splitting is much smaller than all other energy-scales in the problem. The form~\eqref{eqn:N_T_gen} allows distinction of two limits. The first one corresponds to $\mu_0 z_0 \Lambda \lesssim 1$, the regime we call weak superconductivity ($\Lambda \propto n_s \to 0$). In this case, the transverse noise is determined by the normal-fluid contribution:
\beq
\mathcal{N}_{\rm T}(\Omega) \approx  \frac{\mu_0^2 k_B T}{16 \pi z^2_{0}}  \text{Re}\left\{\sigma_n^{\rm T}\Big(\frac{1}{2z_0},\Omega\Big)\right\}
\label{eq:Nz1}.
\eeq
The same expression is known for the simple metallic phase~\cite{Agarwal2017}. The second limit corresponds to $\mu_0 z_0 \Lambda \gtrsim 1$, the regime of strong superconductivity, in which case:
\beq
\mathcal{N}_{\rm T}(\Omega) \approx \frac{3 k_B T}{8 \pi z_0^4\Lambda^2}\text{Re}\left\{\sigma_n^{\rm T} \Big(\frac{3}{2z_0},\Omega\Big) \right\}
\label{eq:Nz2}.
\eeq
In contrast to the metallic behavior~\eqref{eq:Nz1}, the presence of the super-flow gives additional $1/z_0^2$ suppression. Further, from Eqs.~\eqref{eq:Nz1} and \eqref{eq:Nz2}, we note that in both limits, $\N_{\rm T}$ is essentially set by the non-local quasi-static conductivity of the normal fluid $\sigma_n^{\rm T}(q \sim 1/z_0, \Omega \to 0)$. We remark that this reverse order of limits compared to the usual probes such as dc conductivity, where one takes $q \to 0$ first and then $\Omega \to 0$, renders qubit sensors promising to study novel transport regimes, determined by a complicated interplay of various length scales in the superconductor. We also note that the length scale $\mu_0\Lambda$, which is used to separate the weak and strong superconducting regimes, is the well-known ``Pearl length" \cite{pearl1964current}, which is the characteristic length scale associated with the magnetic field distribution around a vortex in a thin-film superconductor.

For the remainder of this section, we focus on calculating the transverse quasiparticle conductivity $\sigma_n^{\rm T}(q \sim 1/z_0, \Omega \to 0)$ in both clean and disordered superconductors, with different pairing symmetries and different spin structures of the superconducting order parameter. We compute $\sigma_n^{\rm T}$ within the linear response formalism using the standard Kubo formula~\cite{altland2010condensed,coleman_2015}, which relates the normal conductivity $\sigma_n^{\rm T}$ to the imaginary-time correlation function of transverse normal currents $j_{\rm T}(\q, \tau) = (\hat{z} \times \hat{\q}) \cdot  \j^n(\q, \tau)$. Specifically, we obtain $\sigma^{\rm T}_n(\q,\Omega)$ from $\Pi_{\rm T}(\q, \tau) = -\frac{1}{{\cal A}} \langle T_\tau (j_{\rm T}(\q, \tau) j_{\rm T}(-\q, 0)) \rangle$ via analytic continuation from imaginary to real frequency:
\beq
\text{Re}[\sigma^{\rm T}_n(\q,\Omega)] =  -\frac{\text{Im}[\Pi_{\rm T}(\q, i\Omega_n \rightarrow \Omega + i 0^+)]}{\Omega},
\eeq
where $\Pi_{\rm T}(\q, i \Omega_n) = \int_{0}^{\beta} d\tau \, e^{i \Omega_n \tau} \Pi_{\rm T}(\q, \tau)$ and ${\cal A}$ is the area of the 2D sample. For simplicity and physical transparency, we evaluate the transverse quasiparticle conductivity within the mean-field BCS theory.  We remark that the BCS Hamiltonian already takes into account the effective short-range attractive interaction between the pairing electrons but neglects the long-range Coulomb repulsion. On the other hand, this long-range interaction is not expected to modify the transverse conductivity because transverse current fluctuations do not perturb local charge density, which experiences strong Coulomb forces. Hence, it is legitimate to evaluate the transverse conductivity to the one-loop level for the Bogoliubov quasiparticles. (In contrast, if one is interested in the longitudinal quasiparticle conductivity, then the long-range Coulomb interaction might play a major role.) Below we focus on presenting the main physical picture and relegate tedious calculations to Appendix~\ref{app:SigmaN}.

\subsection{Singlet superconductors}
The BCS Hamiltonian for singlet superconductors is given in terms of electron operators $c_{\k, \sigma}$, their bare dispersion $\xi_{\k} = \varepsilon_\k - \mu$, and gap-function $\Delta_\k$ as (we assume $\xi_{\k} = \xi_{-\k}$ and work in the gauge with $\Delta_\k \in \mathbf{R}$):
\beq
H_{\rm BCS} = \sum_{\k} \Psi^\dagger_\k h_\k \Psi_{\k}, \quad h_\k = \begin{pmatrix} \xi_\k & \Delta_\k \\ \Delta_\k & -\xi_{\k} \end{pmatrix},
\eeq
where $\Psi_\k = (
c_{\k, \uparrow},\,
c^\dagger_{-\k, \downarrow})^T$ is the Nambu spinor. Within this model, the quasiparticle excitation energy is $E_\k = \sqrt{\xi_\k^2 + \Delta_\k^2}$. Introducing a phenomenological lifetime via an electron self-energy $\Sigma(\k,i\omega_n)$, the Matsubara Green's function (it is a $2\times 2$ matrix in the Nambu space) is given by ($\omega_n = (2n+1)\pi k_B T$)
\beq
G(\k, i \omega_n) &=& (i \omega_n - \Sigma_{\k,i\omega_n} - h_\k)^{-1}.
\eeq
Within a simple model of isotropic disorder scattering, the retarded self-energy $\Sigma^R(\k, \omega)$, obtained from $\Sigma(\k,i\omega_n)$ by analytic continuation to real frequencies, can be approximated as $\Sigma^R(\k, \omega \rightarrow 0) \approx -i \Gamma_0$, where $\Gamma_0$ is simply the isotropic scattering rate of electrons at the Fermi surface (we assume that the real part of $\Sigma^{R}(\k,\omega)$ just renormalizes the bare dispersion). To evaluate the dissipative part of the normal conductivity $\text{Re}[\sigma_n^{\rm T} (\q, \Omega)]$, one needs to consider only the paramagnetic part of the current operator, which is given in terms of the spinor $\Psi_\k$ and quasiparticle velocity $\bm v(\k) = \partial_\k \xi_\k \approx v_F \hat{k}$ (simplifying to a circular Fermi surface) as:
\begin{gather*}
    j_{\alpha}(\q) = e \sum_{\k,\sigma} v_\alpha(\k) c^{\dagger}_{\k_-,\sigma} c_{\k_+, \sigma} = e \sum_{\k} v_\alpha(\k) \psi^{\dagger}_{\k_-} \psi_{\k_+},
\label{eq:J}
\end{gather*}
where $\k_\pm = \k \pm \q/2$. 
As we show in Appendix~\ref{app:SigmaN}, the real part of $\sigma_n^{\rm T}$ can be conveniently written in terms of the spectral function $A(\k, \omega) = - \frac{1}{\pi} \text{Im}[G^R(\k,\omega)]$ as:
\begin{align}
    \sigma_{n}^{\rm T}(\q,\Omega)  =  e^2 \pi  \int & \frac{d^2k}{(2\pi)^2} \int d\omega \; v_T^2 \left(-\frac{\partial n_F(\omega)}{\partial \omega}\right)\notag\\
& \quad \times \Tr[A(\k_-,\omega) A(\k_+, \omega + \Omega)], \label{eq:sigmaN}
\end{align}
Here 
$\bm{v}_{\rm T} = v_F(\hat{q} \times \hat{k})$ is the transverse component of the electron velocity and 
$n_F(\omega) = [\exp(\beta\omega) + 1]^{-1}$ is the Fermi function ($\beta = 1/k_B T$).
Further analytical progress in understanding the transverse quasiparticle conductivity can be achieved by separately considering the clean and disordered limits.

In the clean limit,  $\Gamma_0 \to 0$, the mean-free path $\ell_{\rm MF} = v_F/2 \Gamma_0 \gtrsim \lambda_T$ and the disorder smearing of the spectral function vanishes (the Pauli matrices $\tau^\alpha$ act in the particle-hole/Nambu space):
\begin{equation*}
A(\k,\omega) = \frac{\omega + \xi_\k \tau^z + \Delta_\k \tau^x}{2 E_\k} \left( \delta(\omega - E_\k) - \delta(\omega + E_\k) \right)
\label{eq:clean_SF_main}.
\end{equation*}
In this case, the transverse normal conductivity is dominated by resonant particle-hole excitations across a shell of width $\Omega$, with a relative momentum $\q$ [see Appendix~\ref{app:SigmaN} for additional discussion]:
\begin{align}
    \sigma_{n}^{\rm T}(\q,\Omega) \approx -\frac{e^2}{2\pi} \int d^2k \; v_T^2 \, n_F^\prime(E_\k) \, \delta (\Omega + E_{\k_+} - E_{\k_-}).
\label{eq:sigma_clean}
\end{align}
We note that the contributions to $\sigma_{n}^{\rm T}(\q,\Omega)$ due to simultaneous excitation of two quasiparticles (i.e, for $\Omega = E_{\k_+} + E_{\k_-}$) are suppressed by an additional factor of $q^2$ due to superconducting coherence factors, as discussed in Appendix~\ref{app:SigmaN}.
Accordingly, such two-particle contributions can be neglected not only for fully gapped superconductors (where $\Omega \ll 2 \Delta \leq E_{\k_+} + E_{\k_-}$), but also for nodal superconductors, as long as $q$ is much smaller than the Fermi momentum $k_F$.

In the disordered limit, the spectral function $A(\k,\omega)$ is smeared out by the disorder-induced self-energy $\Gamma_0$, so that $A(\k,\omega)$ acquires a Lorentzian form for $\omega \lesssim k_B T$:
 \begin{equation*}
    A(\k, \omega \lesssim k_B T) \approx \frac{\Gamma_0}{\pi(\Gamma_0^2 + E_\k^2)} \begin{pmatrix} 1 & 0 \\ 0 & 1 \end{pmatrix}.
\end{equation*}
The smooth behavior of the spectral function at small $\omega$ makes it legitimate to approximate the Fermi function derivative $-n_F^\prime(\omega)$ in Eq.~\eqref{eq:sigmaN} by a delta function $\delta(\omega)$, leading to:
\begin{gather}
\sigma_{n}^{\rm T}(\q, \Omega) \approx \frac{e^2}{2\pi^3}  \int d^2k \; v_T^2 \frac{\Gamma_0}{\Gamma_0^2 + E_{\k_+}^2} \frac{\Gamma_0}{\Gamma_0^2 + E_{\k_-}^2} .
\label{eq:sigma_dirty}
\end{gather}

Below we apply the results in Eqs.~\eqref{eq:sigma_clean} and~\eqref{eq:sigma_dirty} to investigate and contrast the properties of s-wave and d-wave superconductors in various regimes.

\subsubsection{s-wave superconductors}

\begin{figure}[t!]
    \centering
    \includegraphics[width=0.45\textwidth]{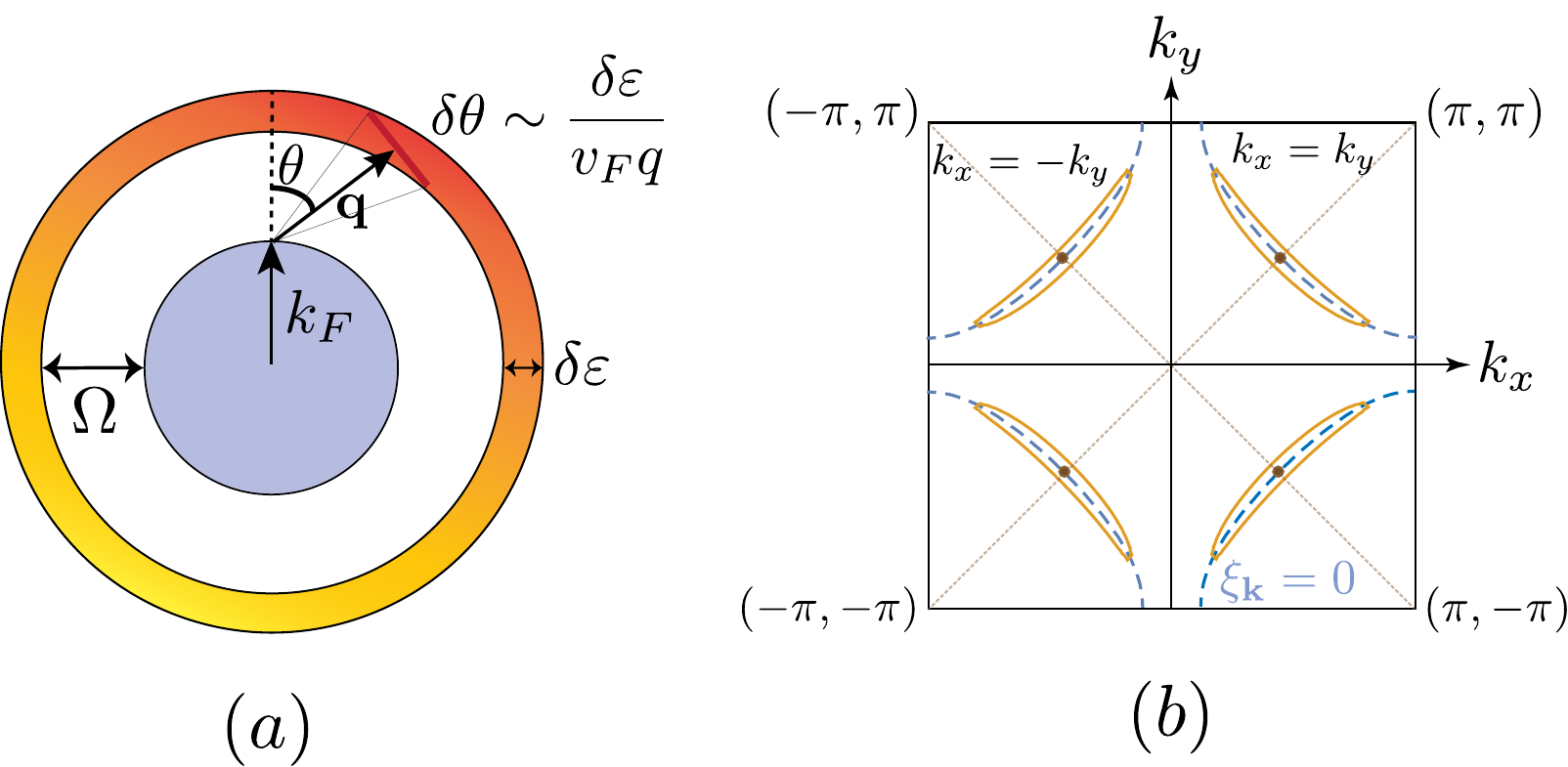}
    \caption{(a) Phase space of the quasiparticle excitations in s-wave superconductors, showing that the angular extent scales as $1/q$. (b) Banana-shaped anisotropic quasi-energy contours (yellow lines) of Dirac cones near the nodal points (brown dots) of a d-wave superconductor. The blue dotted lines denote the underlying Fermi surface $\xi_{\k}=0$.}
    \label{fig:PhaseSpFig}
\end{figure}

We begin by considering the case of clean s-wave superconductors with $\Delta_\k = \Delta$ so that the quasiparticle excitations are gapped and have zero gap velocity. 
Numerical analysis of Eq.~\eqref{eq:sigma_clean}, cf. Fig.~\ref{fig:sigmaRegimesGamma}, indicates that $\sigma_n^{\rm T}(\bm q,\Omega \to 0)$ scales as $1/q$ (up to nonessential logarithmic corrections), both deep in the superconducting phase and near the transition temperature where the superconductivity is suppressed. This behavior can be understood as follows. Approximating $\xi_{\k} \approx v_F(k - k_F)$, we observe that the angular integral over $\k$ in Eq.~\eqref{eq:sigma_clean} is rather restricted for fixed values of $\q$ and $\Omega$. Figure~\ref{fig:PhaseSpFig}(a) shows the phase space of quasiparticle excitations that contribute to the transverse conductivity, where for visibility, we broadened the outer energy circle to have a finite width $\delta \varepsilon$. Specifically, for a given momentum $\bm q$, as follows from the geometry of the Fermi surface, this phase space scales as $1/v_F q$, explaining the behavior of the transverse normal conductivity.

To gain further insight into the properties of $\sigma_n^{\rm T}$, we now focus on the physically relevant limit $\Omega \to 0$ and consider low temperatures first. In this case, we get:
\beq
\sigma_n^{\rm T}(\q, \Omega \to 0) &=& \frac{e^2 v_F k_F}{\pi T q \xi_T} I(\beta \Delta)\label{eq:SigmaSwave},
\eeq
where
\begin{equation}
I(\beta \Delta) = \int_{0}^{\infty} dt \;\frac{ \sqrt{1+ 1/t^{2}} }{4 \cosh^2\left((\beta \Delta /2) \sqrt{t^2 + 1} \right)}. \label{eqn:I_exp}
\end{equation}
This integral contains a weak logarithmic divergence due to the singular quasiparticle density of states $\nu(E) = E/\sqrt{E^2 - \Delta^2}$ near the gap threshold $E\approx \Delta$. In practice, $I(\beta\Delta)$ is regularized by either a small disorder strength or by small $\Omega/q v_F$ (see Appendix \ref{app:SigmaN} for additional discussion). 
Therefore, for realistic experimental parameters, we do not expect this divergence to play a crucial role (see Fig.~\ref{fig:sigmaRegimesGamma}), and we focus on the physically important feature of $I(\beta \Delta)$ --- its temperature dependence.
Deep in the superconducting phase, with $k_B T \lesssim \Delta(T)$, thermal gapped quasiparticle excitations that carry the transverse normal current are suppressed, manifesting as $I(\beta\Delta)\sim \exp(-\beta\Delta)$ in Eq.~\eqref{eqn:I_exp}. Therefore, both the transverse normal conductivity $\sigma_n^{\rm T}$, cf. Eq.~\eqref{eq:SigmaSwave}, and the transverse noise $\N_{\rm T}$ become exponentially suppressed at low temperatures. We conclude that a hallmark of the superconducting phase transition in qubit-based experiments is the exponential suppression of the transverse noise $\N_{\rm T}$ with decreasing temperature below $T_c$.

In the regime of weak superconductivity with small quasiparticle gap $\Delta(T) \lesssim k_B T$, corresponding to temperatures close to $T_c$, the transverse noise displays a different behavior with $T$.
Upon increasing $T$ towards $T_c$, the superconducting coherence length $\xi_T = v_F/\Delta(T)$ increases, while the thermal wavelength $\lambda_T = v_F/k_B T$ decreases. Within the BCS mean-field theory with $\Delta(T) \approx 3 k_B T_c \sqrt{1 - T/T_c}$, these two length scales cross each other near $T \approx 0.8 T_c$. For temperatures above this crossing point, one replaces $\xi_T$ with $\lambda_T$ in Eq.~\eqref{eq:SigmaSwave} and sets $I(\beta \Delta) \approx 1$, thereby obtaining temperature independent normal conductivity $\sigma^{\rm T}_n$, just as in a Fermi liquid \cite{Khoo}:
\beq
\sigma_n^{\rm T}(\q, \Omega \to 0) = \frac{e^2 v_F k_F}{ \pi T q \lambda_T } = \frac{e^2  k_F}{\pi q}. 
\label{eq:SigmaCleanMetal}
\eeq
Accordingly, $\N_{\rm T}\propto T$, reminiscent of the Johnson-Nyquist noise in metals. We remark that Eq.~\eqref{eq:SigmaCleanMetal} might not be entirely correct for a narrow temperature window near $T_c$, where fluctuations effects become essential.

To conclude the discussion of clean s-wave superconductors, by considering $\sigma_n^{\rm T}(q \sim 1/z_0, 0)$, we now examine distance scalings of $\N_{\rm T}$ in both regimes. In the strong superconducting regime, as follows from Eq.~\eqref{eq:Nz2}, we have $\N_{\rm T}\sim 1/z_0^3$, while in the weak superconducting regime, $\N_{\rm T}\sim 1/z_0$, cf. Eq.~\eqref{eq:Nz1}.
This latter behavior can be simply understood by replacing the scattering time $\tau$ in the Drude formula $\sigma = ne^2 \tau/m$ with the time $\tau^\prime = z_0/v_F$, taken by a ballistic quasiparticle at the Fermi surface to travel a linear distance $z_0$ that the qubit can see~\cite{Kolkowitz}. In this argument, we have implicitly used $z_0 \ll \ell_{\rm MF}$, which is valid for clean superconductors with $\ell_{\rm MF} \to \infty$. If $z_0$ is larger than the mean-free path $\ell_{\rm MF}$, the qubit becomes sensitive to multiple scattering events, in which case the conductivity $\sigma_n^{\rm T}$ saturates to a non-singular constant as $q \to 0$. As such, the transverse noise will display $1/z_0^4$ ($1/z_0^2$) scaling in the strong (weak) superconducting regime, as we show next.

We turn to discuss properties of disordered s-wave superconductors and consider first the case of large sample-probe distance $z_0 \gg \ell_{\rm MF}$, equivalent to $q \ell_{\rm MF} \ll 1$.
In this limit, we can explicitly carry out the integral in Eq.~\eqref{eq:sigma_dirty}, with the result (see Appendix~\ref{app:SigmaN} for details):
\beq
\sigma_n^{\rm T}(\q \to 0, 0)  =  \frac{e^2 k_F v_F}{4\pi}\frac{\Gamma_0^2}{(\Gamma_0^2 + \Delta^2)^{3/2}}
\label{eq:sigmaSDis}.
\eeq
We note that Eq.~\eqref{eq:sigmaSDis} reproduces the Drude formula $\sigma = n e^2 \tau/m$, valid in the metallic limit with $\Delta = 0$. 
Here $n =  k_F^2/2 \pi$ is the electron density (including spin),  
$\tau = 1/2 \Gamma_0$ is the electron lifetime, 
and $m = k_F/v_F$ is the effective electron mass. 
The fact that $\sigma_n^{\rm T}(\q \to 0, 0)$ approaches a finite constant explains the mentioned dependence of the transverse noise $\N_{\rm T}$ on the sample-probe distance $z_0$.
On lowering $T$ below $T_c$, the transverse normal conductivity becomes algebraically suppressed with temperature due to the onset of the superconducting gap $\Delta(T)\sim\sqrt{1-T/T_c}$, leading to a corresponding algebraic suppression of $\N_{\rm T}$. 
Deep in the superconducting phase, where $\Delta(T)$ depends weakly on $T$, the transverse conductivity (almost) becomes temperature independent, leading to $\N_{\rm T}\propto T$.

\begin{figure}
    \centering
    \includegraphics[width=0.45\textwidth]{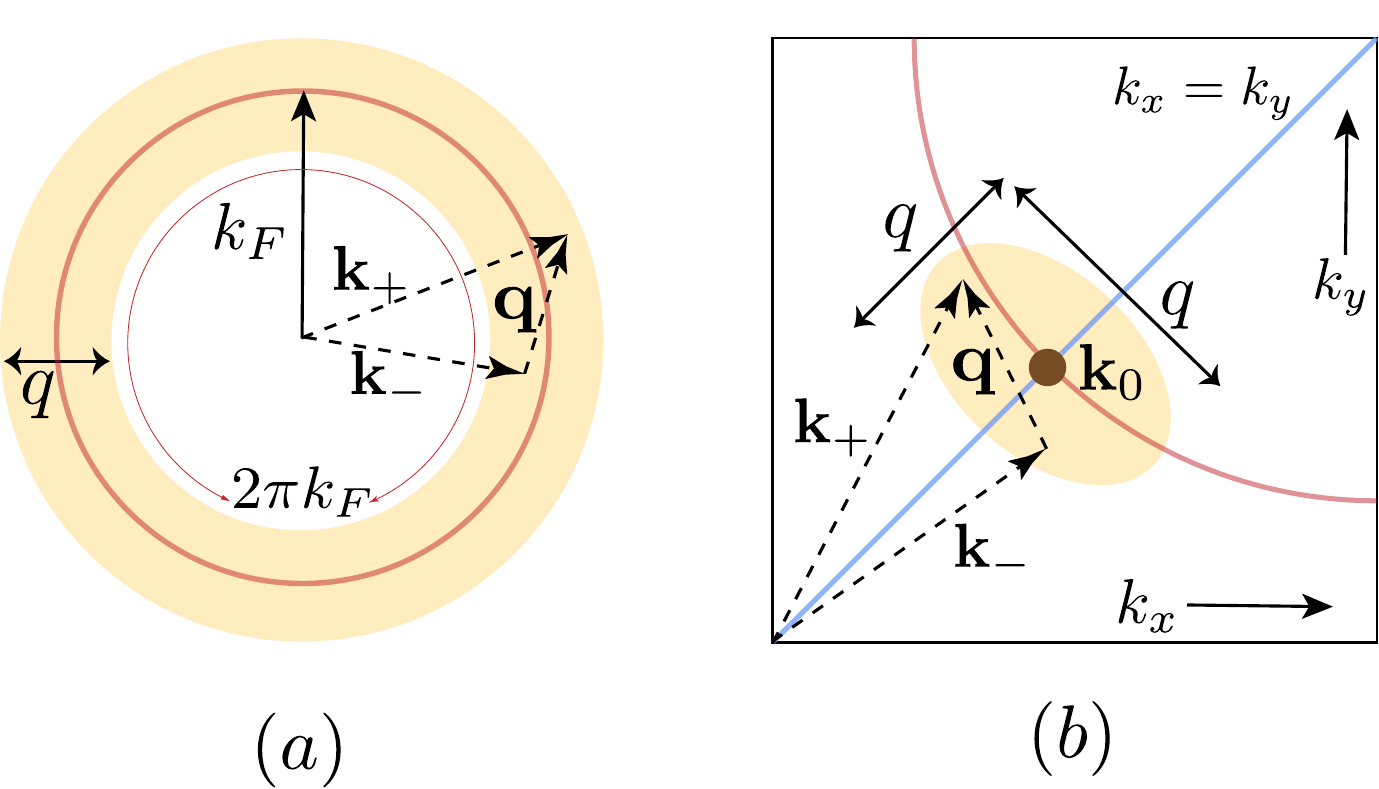}
    \caption{Phase space of the quasiparticle excitations that contribute to $\sigma_n^{\rm T}(q,0)$ in the disordered limit, at small sample-probe distance $q \ell_{\rm MF} \gg 1$. (a) s-wave: $\k_+$ and $\k_-$ lie in an annular strip (shaded yellow) of circumference $2\pi k_F$ and thickness $q$. (b) d-wave: $\k_+$ and $\k_-$ lie in a patch of area $\sim q^2$ around the node $\k_0$ (only a single node/quarter of the first Brillouin Zone is shown).} 
    \label{fig:PhaseSpFigDis}
\end{figure}

Remarkably, by tuning the sample-probe distance $z_0$, qubit-based experiments can gain access to probe the transverse normal conductivity at large momenta. For disordered s-wave superconductors, a new transport regime emerges for $\max \{ \ell_{\rm MF}^{-2}, \xi_T^{-2} \} \lesssim q^2 \ll k_F^2$, where we have 
\beq
\sigma_n^{\rm T}(\q, 0)  \xrightarrow{v_F^2 q^2 \gtrsim \Gamma_0^2 + \Delta^2} \frac{8 e^2 k_F \Gamma_0^2}{\pi^2 v_F^2 q^3}.
\eeq
This result is obtained from both numerical analyses of Eq.~\eqref{eq:sigma_dirty}, cf. Fig.~\ref{fig:sigmaRegimesGamma}, together with analytical calculations outlined in Appendix~\ref{app:SigmaN}. In practice, since $q\sim 1/z_0$, to probe this transport regime, one needs to choose the sample-probe distance $z_0 \lesssim \min\{ \ell_{\rm MF}, \xi_T\}$ to be smaller than both the mean-free path $\ell_{\rm MF}$ and the superconducting coherence length $\xi_T$. The behavior $\sigma_n^{\rm T}(\q,0)\sim 1/q^3$ can be intuitively understood by a careful consideration of the phase space of relevant quasiparticle excitations, similar to our discussion of the clean limit. The momentum integral in Eq.~\eqref{eq:sigma_dirty} is dominated by processes with $|\k| \approx k_F$, so that  both $|\k_+|$ and $|\k_-|$ lie in a momentum window of size $q$ around $k_F$. In this region, the product of spectral functions $A(\k_+, 0) A(\k_-, 0)$ scales as $1/\xi_{\k_+}^2 \xi_{\k_-}^2 \sim 1/(v_F q)^4$. 
The annular strip in momentum space, where this contribution comes from, has width $q$ and circumference $2\pi k_F$, as shown in Fig.~\ref{fig:PhaseSpFigDis}(a). 
This results in an additional factor of $2\pi k_F q$ to the integral, and, therefore, the transverse normal conductivity scales as $2\pi k_F q/(v_F q)^4 \sim 1/q^3$. For the transverse noise, we find that $\N_{\rm T} \sim 1/z_0$ in the regime of strong superconductivity for small qubit-probe distance $z_0$, while it decreases linearly with $z_0$ in the weak superconducting regime.

\subsubsection{d-wave superconductors}

We turn to investigate superconductors with d-wave symmetry of the order parameter. For concreteness, we consider a square lattice and assume $\Delta_\k = \Delta(T) (\cos k_x - \cos k_y)$. The key feature of d-wave superconductors is the presence of gapless quasiparticles, located near the four Dirac points, given by $k_x = \pm k_y$ and $|\k| = k_F$. We note that  the Fermi velocity $\bm v_F = \nabla_\k \xi_\k$ and the gap velocity $\bm v_\Delta = \nabla_\k \Delta_\k$ are orthogonal to each other at each node $\k_0$, which allows us to approximate the quasiparticle energy as $E_\k = \sqrt{v_F^2 k_\parallel^2 + v_\Delta^2 k_\perp^2}$, where $\k - \k_0 = (k_\parallel,k_\perp)$.

We begin by considering the clean limit first, in which case the transverse conductivity is determined by resonant quasiparticle excitations across energy shells of width $\Omega$, cf. Eq.~\eqref{eq:sigma_clean}. These excitations take place on each of the Dirac cones (rather than on the Fermi surface), and geometrical considerations here are identical to our discussion of clean s-wave superconductors. For the d-wave case, we also find $\sigma_n^{\rm T}\sim 1/q$ [see Appendix~\ref{app:SigmaN} for more details]:
\beq
\sigma_n^{\rm T}(\q, \Omega \to 0) = \frac{e^2 v_F \ln(2)}{\beta \pi q v_\Delta}
\label{eq:SigmaDwaveClean}.
\eeq
The key difference compared to the s-wave case is the presence of a non-zero gap velocity $v_\Delta$, which is typically much smaller than the Fermi velocity $v_F$ and leads to very anisotropic (banana-shaped) constant-energy contours (see Fig.~\ref{fig:PhaseSpFig}(b)).

We find that the transverse noise for clean d-wave superconductors is given by:
\beq
\N_{\rm T} & =&  \frac{2\mu_0^2 e^2 \ln(2)}{(2 \pi)^3 \beta^2 z_0 v_\Delta} \left[ K\left( 1- \frac{v_\Delta^2}{v_F^2} \right) +\frac{v_F}{v_\Delta} K\left( 1- \frac{v_F^2}{v_\Delta^2} \right)  \right] \nn &&\xrightarrow{v_\Delta \ll v_F} \frac{4 \mu_0^2 e^2 (k_B T)^2 \ln(2)}{(2 \pi)^3 z_0 v_\Delta} \ln\left( \frac{4 v_F}{v_\Delta}\right),
\label{eq:NoiseDwaveClean}
\eeq
where $K(x) = \int_0^{\pi/2} d\theta (1 - x \sin^2\theta)^{-1/2}$ is the elliptic integral.
The distance-scaling of $\N_{\rm T}$ here is the same as in the s-wave situation: it scales as $1/z_0^3$  ($1/z_0$) in the strong (weak) superconducting regime. The temperature dependence is different: In contrast to the s-wave case with exponentially suppressed transverse noise, here we have $\N_{\rm T}\propto T^2$ for $T \ll T_c$. This is a consequence of the regular power-law density of states of gapless quasiparticles. Another significant difference is the appearance of $v_\Delta^{-1} \ln(4v_F/v_\Delta)$ in Eq.~\eqref{eq:NoiseDwaveClean} which has a two-fold effect.
First, it gives notable enhancement since $v_\Delta/v_F$ is typically small. 
Second, it affects the temperature dependence of the transverse noise close to the critical temperature. 
Since within the mean-field theory $v_\Delta \sim (T - T_c)^{1/2}$ near $T_c$, it gives a sharper increase of the transverse noise for the d-wave case, as $T$ approaches $T_c$ from below. We remark that this apparent divergence for $T\to T_c$ is in practice smoothed out when the gap magnitude $\Delta(T)$ becomes smaller than $k_B T$. In this limit, the description in terms of Dirac cones is no longer appropriate, as quasiparticle excitations all around the Fermi surface start to contribute to conduction.

\renewcommand{\arraystretch}{1.5}
\begin{table}[!t]
    \centering
    \begin{tabular}{|c||c|c|c|c|} \hline
    \multirow{2}{*}{$\sigma_n^{\rm T}(q,0)$} &
    \multicolumn{2}{|c|}{s-wave} & \multicolumn{2}{|c|}{d-wave} \\ 
         & $q \ell_{\rm MF} \ll 1$ &  $q \ell_{\rm MF} \gg 1$ &  $q \ell_{\rm MF} \ll 1$ &   $q \ell_{\rm MF} \gg 1$ \\ \hline \hline
    Clean  & $q^{0}$ & $q^{-1}$ $(^*)$ & $q^{0}$ & $q^{-1}$ \\ \hline
    Disordered & $q^0$ & $q^{-3}$ $(^\dagger)$ & $q^0$ & $q^{-2}$ $(^*)$ \\ \hline
    \end{tabular}
    \caption{ Summary of possible transport regimes, as encoded in the normal-fluid transverse conductivity $\sigma^{\rm T}_n(q,0)$, for s-wave and d-wave superconductors, in clean ($\Gamma_0 \ll k_B T$) and disordered ($\Gamma_0 \gg k_B T$) limits. $(^*)$ implies up to logarithmic corrections, and $(^\dagger)$ also requires $q \xi_T \gg 1$. Compare with the numerical scalings in Fig.~\ref{fig:sigmaRegimesGamma}.}
    \label{tab:sigmaq}
\end{table}

We switch to discuss disordered d-wave superconductors, in which case we obtain [see Appendix~\ref{app:SigmaN} for details]:
\beq
\sigma^{\rm T}_n(\q,\Omega \to 0) = \frac{ e^2 v_F}{\pi^2 v_\Delta}  \frac{ 4\Gamma_0^2 \, \sinh^{-1} \left( \frac{E_\q}{2\Gamma_0} \right) }{E_\q \sqrt{4 \Gamma_0^2 +  E_\q^2 } }, 
\eeq
where $E_\q =\sqrt{ v_F^2 q_\parallel^2 + v_\Delta^2 q_\perp^2}$. When the qubit is placed far from the sample $\ell_{\rm MF} \ll z_0$ (equivalently, $E_\q \ll \Gamma_0$ for $q z_0 \simeq 1$), we recover the {\it universal} Durst-Lee result for the conductivity that is independent of the disorder strength~\cite{Lee1993,DurstLee}:
\beq
\sigma_n^{\rm T}(\q \to 0, 0) = \frac{e^2 v_F}{\pi^2 v_\Delta}.
\label{eq:SigmaNdwave1}
\eeq
We remark that in obtaining this result, we neglected disorder ladder corrections to conductivity, which are expected to be nonzero for d-wave superconductors but not expected to qualitatively affect the conductivity (see Ref.~\onlinecite{DurstLee} for a related discussion). In this regime, the distance scaling of the transverse noise is identical to the s-wave scenario. In the opposite limit corresponding to $z_0 \lesssim \ell_{\rm MF}$, we find that $\sigma_n^{\rm T}(q,0)\sim 1/q^2$ (up to logarithmic corrections):
\beq
\sigma_n^{\rm T}(\q,0) \xrightarrow{q \gtrsim \ell_{\rm MF}^{-1} } \frac{ e^2 v_F}{\pi^2 v_\Delta}  \left( \frac{2\Gamma_0}{E_\q} \right)^2  \, \ln\left( \frac{E_\q}{\Gamma_0} \right). 
\label{eq:SigmaNdwave2}
\eeq
Intuitively, one can again understand this $1/q^2$ scaling by examining the phase space of relevant quasiparticle excitations. For the d-wave case, the integral in Eq.~\eqref{eq:sigma_dirty} is dominated by processes near the Dirac cones, such that $\k_+$ and $\k_-$ lie in a momentum window of size $q$ around each of the nodes $\k_0$. In these regions, the product of spectral functions $A(\k_+, 0) A(\k_-, 0)$ behaves as $1/E_{\k_+}^2 E_{\k_-}^2 \sim 1/q^4$, while the area of the patches, where this contribution comes from, is roughly $q^2$, as shown in Fig.~\ref{fig:PhaseSpFigDis}(b). We, therefore, conclude that the transverse normal conductivity scales as $1/q^4 \times q^2 \sim 1/q^2$, consistent with Eq.~\eqref{eq:SigmaNdwave2} up to logarithms. For the transverse noise, we obtain that $\N_{\rm T}$ is independent of $z_0$ (decays as $1/z_0^2$) in the weak (strong) superconducting regime.

\begin{figure*}
    \centering
    \includegraphics[width=0.65\linewidth]{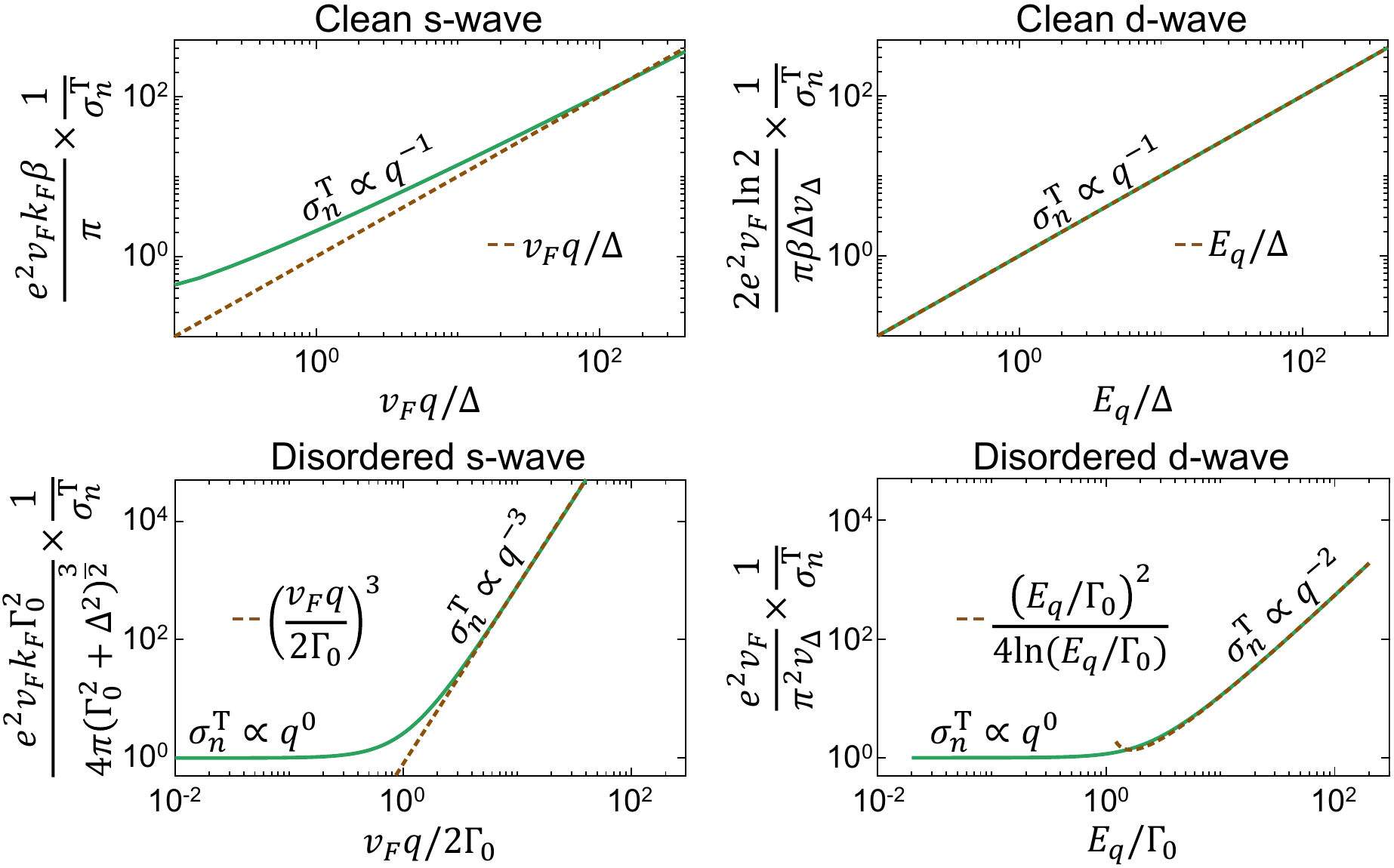}
    \caption{Momentum dependence of $\sigma_n^{\rm T}(\q,\Omega \to 0)$ in various superconductors, obtained by numerical analyses of Eqs.~\eqref{eq:sigma_clean} and~\eqref{eq:sigma_dirty}. These results demonstrate the scalings summarized in Table~\ref{tab:sigmaq}.
    }
    \label{fig:sigmaRegimesGamma}
\end{figure*}

Table \ref{tab:sigmaq} summarizes our findings of possible transport regimes, as encoded in $\sigma_n^{\rm T}(q,0)$, for both clean and disordered superconductors, with different order parameter symmetries. Numerical analyses of Eqs.~\eqref{eq:sigma_clean} and~\eqref{eq:sigma_dirty} are presented in Fig.~\ref{fig:sigmaRegimesGamma}. 
From these, one directly infers the distance-scalings of the transverse noise $\N_{\rm T}$.

\subsection{Triplet superconductors}

We turn to address the question of noise signatures of triplet superconductors in qubit-based experiments, where the order parameter breaks both time-reversal and spin-rotational symmetries. Here we restrict to p-wave superconductivity, corresponding to $\ell = 1$ orbital angular momentum of Cooper pairs, so that the vectorial order parameter $\bm{\Delta}_\k$ links spatial and spin degrees of freedom. The mean-field BCS Hamiltonian can be conveniently expressed in terms of the Balian-Werthamer (BW) spinor~\cite{coleman_2015} $\Psi_\k = (c_\k,\, i \sigma^y c_{-\k}^\dagger)^T$, defined in terms of electron operators $c_{\k} = (c_{\k,\ua},\, c_{\k,\da})^T$ and their time-reversal counterparts, as $H_{\rm BCS} = \sum_{\k \in \frac{1}{2} BZ} \Psi^\dagger_\k h_\k \Psi_\k$ with
\begin{gather}
h_\k = \xi_\k \tau^3 + (\bm \Delta_\k \cdot \bm{\sigma}) \tau^+ + (\bm \Delta_\k^* \cdot \bm{\sigma}) \tau^- .
\end{gather}
Here  $\tau^{\pm} = \frac{1}{2}(\tau^x \pm i \tau^y)$ act in the Nambu (particle-hole) space. 
It is conventional to define the gap function in the spin space as $\bm{\Delta}_\k = \Delta \, \d_\k$, where $\d_\k$ is appropriately normalized over the Fermi surface (assumed to be circular here):
\beq
\int \frac{d\theta_\k}{2\pi} |\d_\k|^2 = 1.
\eeq
Physically, $\d_\k$ denotes the direction normal to the plane of quadrupolar fluctuations of the Cooper pair spin. 
The quasiparticle excitation energy can be shown to be given by $E_{\k,\pm} = \sqrt{\xi_\k^2 + \Delta^2 (|\d_\k|^2 \pm |\d_\k \times \d^*_\k| )}$. 
The normal current operator in terms of the BW spinor reads as
\begin{equation*}
j_{\alpha}(\q) = e \sum_{\k \in \frac{1}{2}BZ} v_\alpha(\k) \Psi^\dagger_{\k - \q/2} \Psi_{\k + \q/2}. 
\end{equation*}
Given a pairing function $\bm{\Delta}_\k$, we use Eq.~\eqref{eq:sigmaN} to evaluate the transverse quasiparticle conductivity $\sigma_n^{\rm T}(\q, \Omega)$ for triplet superconductors, again assuming isotropic disorder scattering.

Here we consider the case of unitary pairing functions, corresponding to $\d_\k $ and $\d^*_\k$ being parallel to each other. Specifically, we choose the form of $\d_\k$ to be analogous to either Balian-Werthamer (BW) or Anderson-Brinkman-Morel (ABM) phases~\cite{He3RMP} of superfluid He$^3$, except in two spatial dimensions. The latter might be relevant~\cite{Maeno2001}, for instance, to Sr$_2$RuO$_4$. Our explicit calculations in Appendix~\ref{app:SigmaN} indicate that for both kinds of pairing functions, the transverse normal conductivity reduces to that of fully gapped isotropic s-wave superconductors, considered above. This conclusion holds for both clean and disordered limits. Even though the transverse current correlation function for p-wave superconductors behaves similarly to the s-wave case, the nature of spin fluctuations can distinguish singlet and triplet pairings, as we discuss in Sec. \ref{sec:SpinN}.

\section{ Noise in Longitudinal sector }
\label{sec:NL}

The results of the previous section show that the presence of superconductivity suppresses the noise due to transverse current fluctuations at low temperatures. Does this mean that longitudinal current fluctuations, neglected above, can start to dominate? 
To address this question, we study coupled dynamics of the order parameter and the electromagnetic field, which allows us to determine the spectrum of longitudinal collective modes and their contribution to $\N_{\rm L}$. Below we investigate both monolayer and bilayer geometries.

\subsection{Longitudinal conductivity and collective modes in monolayers}

We describe spontaneous symmetry breaking via the Ginzburg-Landau free energy:
\begin{align}
    {\cal F}[\psi] =  d \int d^2 {\bm r} & \Big[ \frac{1}{2m^*} \left| \Big(-i \hbar \nabla - e^*  {\cal \bm A} \Big) \psi \right|^2 \notag\\
    &\qquad\qquad\qquad
    + \alpha |\psi|^2 + \frac{\beta}{2} |\psi|^4
    \Big],
\end{align}
where $d$ is the thickness of the film (along this thickness the superconducting order parameter $\psi$ remains homogeneous), 
${\cal \bm A}(\bm r)$ is the in-plane vector potential at $z = 0$, and $m^*$ is the effective Cooper pair mass.
We assume overdamped order-parameter dynamics, captured by a time-dependent Ginzburg Landau (GL) equation \cite{Tinkham}:
\begin{align}
    \tau \left[\frac{\partial}{\partial t} + i e^*  (\delta \mu + \phi) \right] \psi({\bm r},t) =  -  \frac{\delta {\cal F}}{ d \delta \psi^*({\bm r},t)}, 
\end{align}
where $\tau$ is a dimensionless parameter characterizing the order parameter relaxation time. $\phi$ is the scalar potential, and we choose the gauge where $\phi = 0$. $\delta \mu = \chi^{-1} \rho$ is the electrochemical potential describing the coupling between the order parameter and charge fluctuations ($\rho$ is the two-dimensional charge density). $\chi^{-1}$ is the inverse compressibility, a phenomenological parameter in our approach. The two-dimensional superconducting current density reads as
\begin{gather}
    {\bm j}_s = \frac{  d e^*}{2 m^*} \psi^* \Big(-i\hbar \nabla - e^* {\cal \bm A}\Big) \psi + \textrm{c.c.} \label{eqn:j_s}
\end{gather}
We also have the current density due to quasiparticles, which we write as:
\begin{align}
    \bm j_n(\bm q,\omega) = \sigma_n^{\rm L}(\bm q,\omega) ( \mathcal{\bm E}(\bm q,\omega) - i\bm q \delta \mu(\bm q,\omega)),
\end{align}
where ${\cal \bm E}$ is the in-plane electric field at $z =0$, $\bm q$ is the in-plane momentum. 
The conservation of charge is expressed in the continuity equation:
\begin{align}
    \partial_t \rho + \nabla\cdot(\bm j_s + \bm j_n) = 0.
\end{align}

We turn to linearize the above equations of motion on top of the equilibrium state, which has a homogeneous order parameter expectation value $\psi_0 = \sqrt{-\alpha/\beta}$. The dynamics of the order parameter amplitude decouples from the rest of the system and turns out to be overdamped. For this reason, we focus on the dynamics of the order parameter phase $\theta$. The linearized supercurrent reads as
\begin{align}
    \bm j_s = \Lambda \Big( \frac{\Phi_0}{2\pi}\nabla \theta -  {\cal  \bm A} \Big),
\end{align}
where $\Phi_0 = h/2e$ is the magnetic flux quantum. The total longitudinal current density is the sum of the normal and superfluid contributions:
\begin{align}
    j_{\rm L}(\bm q,\omega) = \sigma_n^{\rm L} \Big( i\omega {\cal A}_\parallel - \frac{i q \rho}{\chi} \Big)  + \Lambda \Big( \frac{\Phi_0}{2\pi}i q \theta -  {\cal   A}_\parallel \Big),
\label{eq:JLtot} 
\end{align}
The linearized GL equation for the phase reads as
\begin{align}
    -i\omega \theta + \frac{e^*}{\chi} \rho = -\frac{\hbar^2 \bar{\Gamma}}{2 m^*} \Big(q^2 \theta + \frac{2\pi i}{\Phi_0} q {\cal A}_\parallel\Big), \label{eqn:dyn_phase}
\end{align}
where $\bar{\Gamma} = \tau^{-1}$. By using Eqs.~\eqref{eq:JLtot} and~\eqref{eqn:dyn_phase}, together with the continuity equation $\omega\rho = qj_{\rm L}$, one can (numerically) compute the full longitudinal conductivity. We note that the terms with compressibility $\chi^{-1}$ start to play a role only for $k \gtrsim k_{\rm TF}$, where $k_{\rm TF}$ is the Thomas Fermi screening momentum~\cite{Tinkham}. For a non-interacting two-dimensional Fermi gas, we estimate $k_{\rm TF} \sim a_0^{-1}$, where $a_0$ is the Bohr radius. Since this is a large momentum scale, we now specialize on the case $\chi^{-1} = 0$, i.e. we focus on low momenta $k\lesssim k_{\rm TF}$, in which case, the net longitudinal conductivity can be calculated analytically:
\begin{align}
    \sigma^{\rm L}(\bm q, \omega) =  \sigma_n^{\rm L}(\bm q, \omega) - \frac{\Lambda}{i\omega -  \frac{\bar{\Gamma} \hbar^2 q^2}{2 m^*}}.
\label{eq:SigmaL}
\end{align}

To derive the spectrum of collective modes, we also need to consider the dynamics of the electromagnetic field via Maxwell's equations (we assume that $\mu = \mu_0$):
\begin{gather}
    \epsilon \nabla \cdot {\bm E} = \frac{\rho}{\epsilon_0} \delta(z),\label{eqn:M1}\\
     \nabla \times \bm B = \frac{\epsilon}{c^2}\frac{\partial\bm E}{\partial t} + \mu_0 (\bm j_{s} + \bm j_{n}) \delta(z).\label{eqn:M2}
\end{gather}
In the gauge with zero scalar potential, one writes: $\bm B = \nabla \times \bm A$ and $\bm E = -\partial_t \bm A$. 
For a thin sample, Eqs.~\eqref{eqn:M1} and~\eqref{eqn:M2} can be represented as an interface problem with the following boundary conditions ($+$ ($-$) refers to the top (bottom) boundary):
\begin{align}
    & \epsilon^+ E_{z}^+ - \epsilon^- E_{z}^- = \frac{ \rho}{\epsilon_0}, & B_{z}^+ = B_{z}^-,\\
    & \hat{z}\times (\bm B^+ - \bm B^-) = \mu_0 \bm j, & \bm  E_{t}^+ = \bm E_{t}^-,
\end{align}
where $\bm  E_t = E_x \hat{x} + E_y \hat{y}$ is the tangential component of the electic field and $\epsilon^{\pm}$ are the dielectric constants of the media just above and below the $xy$ plane. 
Equivalently, the boundary conditions can be written solely in terms of the vector potential:
\begin{align*}
    & A_\parallel^+ = A_\parallel^-, ~~
    \partial_z A_\parallel^+ -\partial_z A_\parallel^- = - \mu_0 j_{\rm L} + iq (A_z^+ - A_z^-) 
    \\
    & A_\perp^+ = A_\perp^-, ~~ \partial_z A_\perp^+ -\partial_z A_\perp^- = - \mu_0 j_{\rm T},\\
    & \epsilon^+ A_z^+ - \epsilon^- A_z^- = \frac{\rho}{i \epsilon_0 \omega},
\end{align*}
where we decomposed $\bm A(\bm q, z) = A_\parallel \hat{\bm q} + A_\perp \hat{\bm q} \times \hat{\bm z} + A_z \hat{\bm z}$.

In this system, a collective excitation represents a mode that couples three-dimensional fluctuations of light to two-dimensional fluctuations of the order parameter phase. 
We anticipate such a mode to be an evanescent wave:
\begin{align}
    \bm A (\bm q, z;\omega) = \begin{cases} 
    \bm A_+(\bm q,\omega) e^{-\varkappa z}, & z> 0\\
    \bm A_-(\bm q,\omega) e^{\varkappa z}, & z < 0
    \end{cases}
\end{align}
where $\varkappa = \sqrt{q^2 - \epsilon\omega^2/c^2}$. By solving the Maxwell equations, we obtain:
\begin{align}
    {\cal A}_\parallel(\bm q,\omega) = -\frac{\varkappa}{2 \epsilon \epsilon_0 \omega^2} j_{\rm L}(\bm q,\omega). \label{eqn:A_par_res}
\end{align}
Provided one knows the longitudinal conductivity, the spectrum of the longitudinal collective modes is then defined through:
\begin{align}
1 + \frac{i\varkappa}{2\epsilon \epsilon_0 \omega}\sigma^{\rm L}(q,\omega) = 0,
\label{eqn:long_spect}
\end{align}
which we note is nothing but the usual condition of the vanishing of the (2D) dielectric function, $\epsilon_{\rm L}(q,\omega) = 0$. Anticipating the development of gapless plasmons, let us now focus on low frequencies and low momenta, where one can substitute $\varkappa \approx q$ (we will ignore the light cone, which starts to play a role at negligibly small momenta $\sim\omega/c$). In this regime, as it follows from Eq.~\eqref{eqn:long_spect}, by expanding in powers of $q$,
one can replace $\sigma^{\rm L}(q,\omega)$ with $\sigma^{\rm L}(q = 0,\omega) = \sigma^{\rm T}(q = 0,\omega) = \sigma_n - \frac{\Lambda}{i \omega}$. Interestingly, this argument is generic and does not require explicit computation of the longitudinal conductivity, i.e. one only needs the conductivity at $q=0$, which is the same for both longitudinal and transverse cases. This result is consistent with Eq.~\eqref{eq:SigmaL} at $q = 0$. From Eq.~\eqref{eqn:long_spect}, we obtain the dispersion of the longitudinal collective modes at small momenta $q$:
\begin{align}
\omega_\parallel (q) =   \sqrt{\frac{q  \Lambda}{2\epsilon \epsilon_0} - \Big( \frac{q \sigma_n}{4\epsilon \epsilon_0}  \Big)^2} - i \frac{ q \sigma_n}{4\epsilon \epsilon_0} . \label{eqn:mono_disp}
\end{align}
We note that the primary role of $\sigma_n$ is to provide damping and redshift the otherwise coherent gapless plasmon excitation.

\begin{figure}[t!]
\centering
\includegraphics[width=1\linewidth]{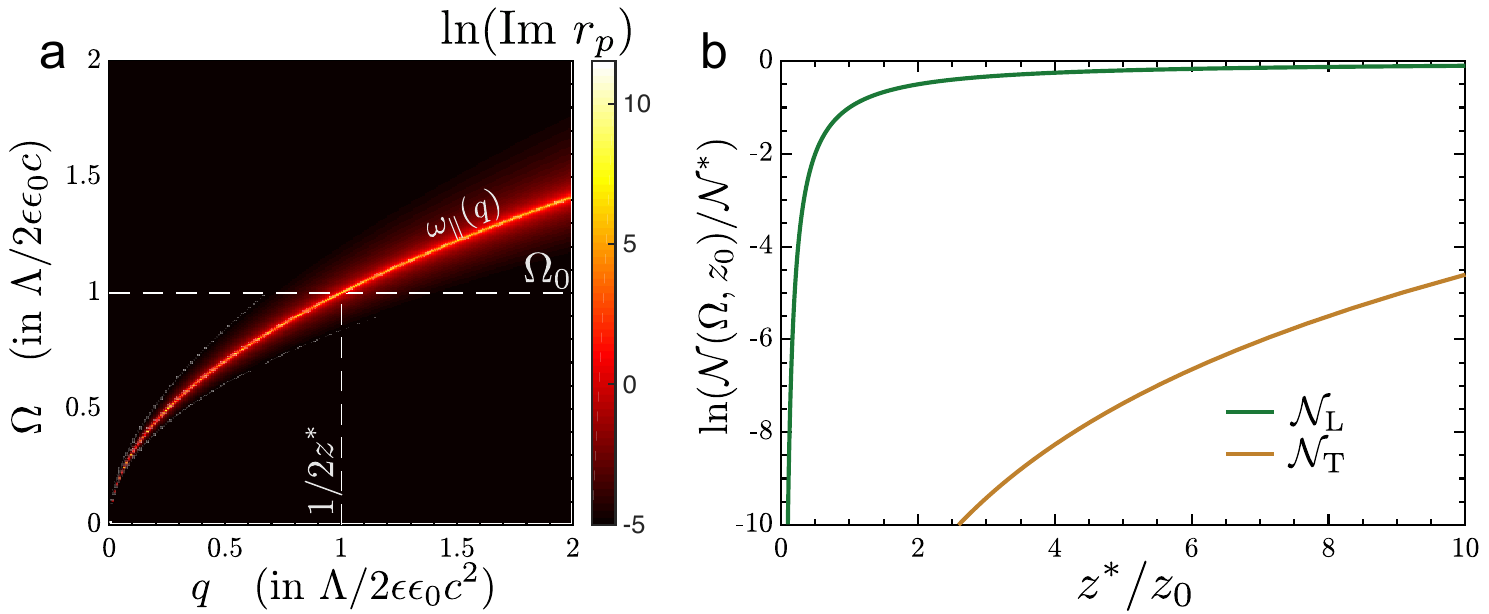} 
\caption{ Noise in monolayers. (a) Reflection coefficient $\text{Im}( r_p(q,q_z,\Omega))$, evaluated for evanescent waves with $q_z\approx iq$, is resonantly enhanced when crossing the plasmon dispersion $\omega_\parallel(q)$, cf. Eq.~\eqref{eqn:mono_disp}. (b) In a monolayer geometry at low temperatures, the transverse noise ${\cal N}_{\rm T}(\Omega,z_{0})$ is suppressed compared to ${\cal N}_{\rm L}(\Omega,z_{0})$. More specifically, for $z_{0} \gg z_0^*(\Omega)$ ($z_0^*$ is defined in the main text), the longitudinal collective modes do not contribute, which also results in suppressed ${\cal N}_{\rm L}$. Once $z_{0} \lesssim z_0^*$, the plasmon branch starts to contribute, and the noise ${\cal N}_{\rm L}(\Omega,z_{0})$ shows a quick saturation at a finite value. At even shorter distances, depending on parameters, ${\cal N}_{\rm T}(\Omega,z_{0})$ can develop and overcome ${\cal N}_{\rm L}(\Omega,z_{0})$.
}
\label{fig::Longitudinal}
\end{figure}

\subsection{Longitudinal noise in monolayers}

We now use the conductivity in Eq.~\eqref{eq:SigmaL} to estimate the longitudinal noise $\N_{\rm L}$ via Eqs.~\eqref{eq:NL} and \eqref{eq:rp}.
Importantly, we observe that for evanescent modes with $q_z = i q$, the reflection coefficient $r_p(\q,\Omega)$ can become resonantly enhanced upon crossing the longitudinal spectrum, cf. Eq.~\eqref{eqn:long_spect}. 
To clarify this point, let us compute $r_p(\q,\Omega)$ for $\bar{\Gamma} = 0$, since this term is relevant only at large momenta, or equivalently at short distances $z_{0}$:
\begin{align}
\text{Im}\{ r_p (\q,\Omega) \} & = \frac{(2 \epsilon \epsilon_0 \Omega^2) (q \Omega \sigma_n^{\rm L})}{ (2 \epsilon \epsilon_0 \Omega^2 -q \Lambda)^2 + (q \Omega \sigma_n^{\rm L})^2} \notag\\
& \approx  \pi \Omega^2 \, \delta\left( \Omega^2 - \frac{q \Lambda} {2\epsilon \epsilon_0}\right) ,
\end{align}
where in the last step, we assumed that the longitudinal quasiparticle conductivity is suppressed at low temperatures. 
This resonant enhancement is clearly visible in a numerical evaluation of $\text{Im}\{ r_p (\q,\Omega) \}$ in Fig.~\ref{fig::Longitudinal}(a).
Therefore, the noise in this limit is given by:
\begin{align}
\mathcal{N}_{\rm L}(\Omega) \approx  \frac{ k_B T \epsilon^2 \Omega^3}{4 \Lambda c^4} \exp\left(-\frac{4 \epsilon \epsilon_0 z_0 \Omega^2}{\Lambda}\right) \nn 
\equiv \mathcal{N}^*(\Omega) \exp\left(-\frac{4 \epsilon \epsilon_0 z_0 \Omega^2}{\Lambda}\right),
\end{align}
where $\mathcal{N}^*(\Omega) =  k_B T \epsilon^2 \Omega^3/4 \Lambda c^4$.
While at low temperatures when $\sigma_n^{\rm L}$ is (exponentially) small, the longitudinal noise can be finite due to collective plasmon modes. 
We demonstrate this in Fig.~\ref{fig::Longitudinal}(b) by plotting the ratio $\mathcal{N}_{\rm L}(\Omega)/\mathcal{N}^*(\Omega)$ as a function of 1/$z_0$.
If we tune the sample-probe distance $z_0$ at a fixed $\Omega$, this ratio is suppressed for large qubit distances. 
It subsequently saturates to a finite value upon crossing the plasmon branch, i.e, for $z_{0}\lesssim z_{0}^*$, where $z_0^*$ is defined via $\Omega = \omega_\parallel(q = 1/2z_{0}^*)$, as shown in Fig.~\ref{fig::Longitudinal}(a).
As pointed out in Ref.~\onlinecite{Shubhayu1}, longitudinal noise can be enhanced by considering a high-$\epsilon$ encapsulating materials such as SrTiO$_3$~\cite{Veyrat781}. Another possible route to enhancing the signal is to increase the probe frequency, $\Omega$. As the noise varies with the cube of $\Omega$, even a modest increase yields a significant enhancement. The magnetic fields corresponding to heightened frequencies may become large, say, on the order of a few Tesla, but destruction of superconductivity can be avoided by orienting the field in the plane of the 2D material, so long as the field strength remains below the Pauli limit.

\subsection{Two-fluid model in bilayers}
A special feature of a bilayer geometry is the Josephson coupling between the layers which leads to the development of two distinct longitudinal collective modes: (i) a symmetric mode that arises from in-phase oscillations of the charge density --- this mode is gapless and closely resembles monolayer plasmons studied above; (ii) an antisymmetric mode that arises from out-of-phase oscillations of the charge density --- this mode is gapped. We anticipate that this gap size is significantly lower compared to $\Delta$, giving the possibility to detect this level splitting at low temperatures with impurity qubits. Below, we derive the spectrum of longitudinal collective modes in bilayer superconductors, and compute their contribution to the longitudinal noise.

As for the monolayer case, we disregard the fluctuations of the order parameter amplitude: this is justified either due to the choice of the overdamped order-parameter dynamics, as above, or, more broadly, in the low-frequency limit $\Omega \ll \Delta$. We then write the free energy only for the order parameter phases and adopt notations commonly used to describe layered superconductors~\cite{PhysRevB.46.366,PhysRevB.50.12831,Koshelev2013}:
\begin{align}
    {\cal F} & = \frac{E_0}{2} \int d^2 {\bm r} \Big[ \Big(\nabla \theta_1 - \frac{2\pi}{\Phi_0}{\cal A}_1\Big)^2 \label{eqn:F_bl}\\
    & + \Big(\nabla \theta_2 - \frac{2\pi}{\Phi_0}{\cal A}_2\Big)^2
     + \frac{2}{\lambda_J^2} [1 - \cos(\theta_1-\theta_2-\vartheta)] \Big].\notag
\end{align}
In Eq.~\eqref{eqn:F_bl}, $\theta_i$ are the order parameter phases in the two layers assumed to be at $z = \pm l/2$. $\mathcal{A}_i$ are the corresponding in-plane projections of the vector potential $\A(\r, z)$. The order parameter phase stiffness $E_0$ is given by:
\beq
E_0 = \frac{\Phi_0^2 l}{4 \pi^2 \mu_0 \lambda_{ab}^2} = \left( \frac{\Phi_0}{2\pi} \right)^2 \Lambda,
\eeq
where $\lambda_{ab}$ is the penetration depth for in-plane currents, and $l$ is the spacing between the layers.
We also defined the Josephson length $\lambda_J = \gamma l$, where the anisotropy parameter $\gamma = \lambda_c/\lambda_{ab} \gg 1$ captures the weak coupling between the layers.
Finally, we have
\begin{align}
    \vartheta = \frac{e^*}{\hbar}\int\limits_{-l/2}^{l/2}dz \, A_z,
\end{align}
so that the free energy in Eq.~\eqref{eqn:F_bl} is gauge invariant. The supercurrent densities can be derived by varying the action with respect to the vector potential. 
The in-plane two-dimensional current-density is given by:
\begin{align}
    \bm j_{\rm{L},i = 1,2} = \Lambda \Big( \frac{\Phi_0}{2\pi}\nabla\theta_i - {\cal \bm A}_i\Big) + \sigma_n^{\rm L}\Big(\mathcal{\bm E}_i - \frac{\nabla \rho_i}{\chi}\Big). \label{eqn:in_plane_j}
\end{align}
In addition to in-plane currents, we also have a three-dimensional current density between the layers, which can again be written as a sum of superfluid and normal-fluid contributions:
\begin{align}
    J_z = J_0 \sin (\theta_1 -\theta_2 - \vartheta) + \frac{1}{\rho_c}\int\limits_{-l/2}^{l/2} \frac{dz}{l} E_z, \label{eqn:J_z}
\end{align}
where $J_0 = \Phi_0 \Lambda/2 \pi \lambda_J^2$ represents the Josephson coupling between the layers, while the second term describes the quasiparticle contribution to interlayer cuurent. 
The latter is expected to be suppressed, both due to anisotropy effects in layered superconductors (typically, $\rho_c \gg \rho_{ab}$) and due to the quasiparticle excitation gap. We assume the current density~\eqref{eqn:J_z} to arise only from quantum tunneling events between the two layers and allow for charge fluctuations to occur only in the superconducting films. Because of the coupling between the layers, the charge conservation is now expressed as:
\begin{align}
    \partial_t \rho_1 + \nabla \cdot \bm j_1 - J_z = 0,\label{eqn:cnt_1}\\
    \partial_t \rho_2 + \nabla \cdot \bm j_2 + J_z = 0.\label{eqn:cnt_2}
\end{align}

\begin{figure}[t!]
\centering
\includegraphics[width=1\linewidth]{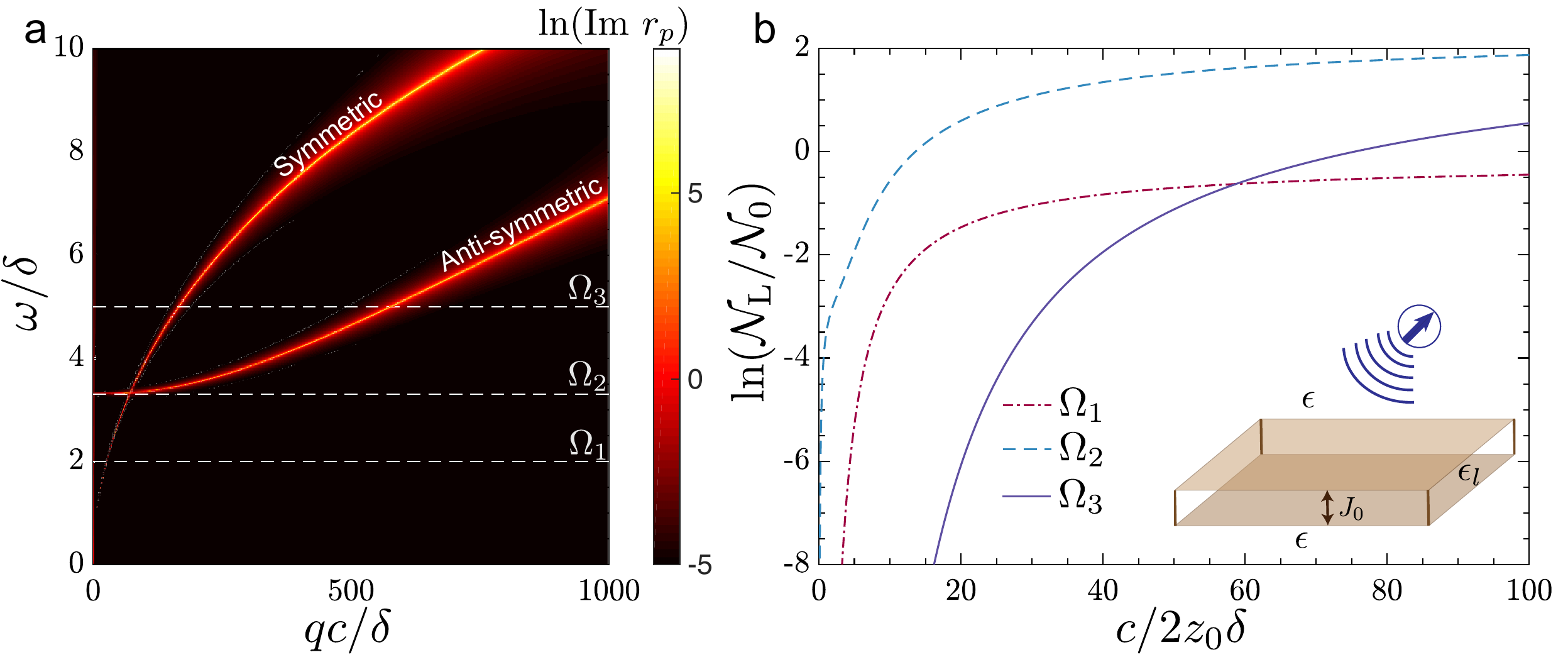} 
\caption{Noise in bilayers. (a) Reflection coefficient $\text{Im}(r_p)$, as a function of $q$ and $\omega$, gets resonantly enhanced upon crossing the longitudinal collective modes. For bilayer, one observes symmetric branch, similar to the monolayer plasmon, and antisymmetric branch, which exhibits opening of the Josephson gap due to both the Josephson coupling between the two layers and the interlayer Coulomb interaction. 
(b) Longitudinal noise ${\cal N}_{\rm L}/{\cal N}_0$ as a function of the qubit distance $z_0$ for the three cuts in (a). Here ${\cal N}_0 = \epsilon\mu_0 k_B T/8\pi c\lambda_{ab}^2$. The middle cut near the Josephson gap indicates that one can detect the antisymmetric branch via qubit sensors. Inset: schematic of bilayer setup. Parameters used: $\epsilon = \epsilon_l = 1$, $\gamma = 5$, $l = 0.01 \lambda_{ab}$, $\chi = 10^3/2\pi\lambda_{ab}$ (the values of other parameters are negligible).}
\label{fig::Im_rp_bilayer}
\end{figure}

To investigate collective modes in a bilayer, we need to consider dynamics of both the order parameter phases and the electromagnetic field that couples to them.
Similarly to the monolayer case, we assume the dynamics of the phases to be overdamped:
\beq
\partial_t \theta_1 + \frac{e^*}{\chi} \rho_1 & = & - \Gamma E_0 \Big[\Big( -\nabla^2 \theta_1 + \frac{2\pi}{\Phi_0} \nabla \cdot {\cal A}_1\Big) \notag\\
    &&\qquad\qquad + \lambda_J^{-2} \sin(\theta_1 - \theta_2 - \vartheta)\Big],\\
    \partial_t \theta_2 + \frac{e^*}{\chi}  \rho_2 & = &  - \Gamma E_0 \Big[\Big( -\nabla^2 \theta_2 + \frac{2\pi}{\Phi_0} \nabla \cdot {\cal A}_2\Big) \notag\\
    &&\qquad\qquad - \lambda_J^{-2} \sin(\theta_1 - \theta_2 - \vartheta)\Big],
\label{eq:tdlg_bl}
\eeq
where the damping coefficient $\Gamma$ is related to the monolayer $\bar{\Gamma}$ by $\bar{\Gamma} = 2 \Gamma |\psi_0|^2$. The dynamics of the electromagnetic field is governed by the Maxwell equations:
\beq
    \epsilon \nabla \cdot {\bm E} &=& \frac{\rho_{1}(\bm r)}{\epsilon_0} \, \delta\Big(z - \frac{l}{2}\Big) +\frac{\rho_{2}(\bm r)}{\epsilon_0} \, \delta\Big(z + \frac{l}{2}\Big),\\
    \nabla \times \bm B &=& \mu_0 \Big[\bm j_{1}(\bm r) \delta\Big(z - \frac{l}{2}\Big) + \bm j_{2}(\bm r) \delta\Big(z + \frac{l}{2}\Big) + \hat{z}J_z f(z) \Big] \notag\\
    && + \epsilon \mu_0  \frac{\partial \E}{\partial t},
\eeq
where $f(z) = 1$ for $-\frac{l}{2} < z < \frac{l}{2}$ and zero otherwise. To deal with the generic scenario, we have allowed for the dielectric constant of the outside medium $\epsilon$ to differ from the dielectric constant $\epsilon_l$ of the material in between the layers [inset of Fig.~\ref{fig::Im_rp_bilayer}(b)]. 
To obtain collective modes, we need to find solutions to the coupled dynamics, which we turn to discuss next.

Like in the monolayer case, a longitudinal collective mode represents an evanescent wave. The reflection symmetry about the plane $z=0$ allows us to decompose the solutions into symmetric and antisymmetric modes. The calculation of the spectrum of the longitudinal collective modes in a bilayer is similar to the one for a monolayer. For this reason, we relegate the details to Appendix~\ref{app:CM_bilayer}, and outline the main results here. We find that the symmetric mode is insensitive to the Josephson coupling between the layers. It is gapless and resembles the monolayer plasmons:
\beq
\omega_s (q) \approx \sqrt{\frac{q  \Lambda}{\epsilon \epsilon_0} - \Big( \frac{q \sigma_n^{\rm L}}{2\epsilon \epsilon_0}  \Big)^2} - i\left(  \frac{ q \sigma_n^{\rm L}}{2\epsilon \epsilon_0} \right). \label{eqn:plasmon_bl}
\eeq
In contrast, the anti-symmetric mode is gapped:
\begin{align}
    \omega_a(q = 0) =  \sqrt{ \delta^2  + \frac{\omega_{ab}^2}{\epsilon_l \gamma^2} - \left(\frac{\Gamma E_0}{\lambda_J^2}\right)^2 } - i\left(\frac{\Gamma E_0}{\lambda_J^2}\right),
\end{align}
where $\delta  = \sqrt{2 e^* J_0/\chi}$ arises due to the Josephson coupling between the layers, while the second term ($\omega_{ab} = c/\lambda_{ab}$) originates from the interlayer Coulomb interaction that penalizes an imbalance of charge.
We note that the antisymmetric mode represents a coherent excitation at low momenta as long as the damping $\Gamma$ is small enough. The spectra of both symmetric and antisymmetric modes are illustrated in Fig.~\ref{fig::Im_rp_bilayer}(a). We remark that the presented framework, which is based on the time-dependent Ginzburg-Landau formalism, is not expected to be quantitatively correct at low temperatures [for a related discussion of low-temperature plasmons in layered superconductors, see Ref.~\onlinecite{PhysRevB.50.12831}]. However, our results qualitatively agree with a more microscopic calculation of the low-energy collective modes of Ref.~\onlinecite{Sun}. In the following subsection, we discuss the impact of the longitudinal collective modes on the magnetic noise.

\subsection{Longitudinal noise in bilayers}

As it follows from Eq.~\eqref{eq:NL}, to obtain the 
longitudinal noise, one needs to evaluate the reflection coefficient $r_p(q,q_z,\omega)$ of the $p$-polarized waves. 
To this end, we solve below the scattering problem for the bilayer:
\beq
\B_{\rm in} &=&  B_0 (\hat{z} \times \hat{q}) e^{i \q \cdot \r - i q^\epsilon_z (z - l/2)} \nn
\B_{\rm r} &=& r_p B_0 (\hat{z} \times \hat{q}) e^{i \q \cdot \r - i q^\epsilon_z (z - l/2)} \nn
\B_{\rm t} &=& t_p B_0 (\hat{z} \times \hat{q}) e^{i \q \cdot \r - i q^\epsilon_z (z + l/2)},
\eeq
where $\B_{\rm in}$, $\B_{\rm r}$, and $\B_{\rm t}$ denote incoming, reflected, and transmitted waves, respectively [see Fig.~\ref{fig:Polarizations} (right panel)]. Here $B_0$ is the amplitude of the incoming wave, assumed to be small. $t_p(\q,q_z,\omega)$ is the transmission coefficient. Since the magnetic noise is essentially determined by the evanescent waves, we substitute $q^\epsilon_z = \sqrt{\epsilon \omega^2/c^2 - q^2} \approx i q$. The evaluation of $r_p$ is analogous to our analysis of the longitudinal collective modes, and we provide the computational details in Appendix \ref{app:CM_bilayer}. Figure~\ref{fig::Im_rp_bilayer}(a) shows the final result for $r_p(q,q_z = iq, \omega)$. Similar to monolayers, we find that this quantity gets resonantly enhanced upon crossing of either of the two longitudinal collective modes.

We now discuss how qubit sensors can be used to probe the antisymmetric mode in bilayer systems.
To this end, we consider three cuts corresponding to fixed qubit frequency, shown in Fig.~\ref{fig::Im_rp_bilayer}(a). For the lowest frequency cut below the Josephson gap, the qubit response is similar to the monolayer case: upon crossing the symmetric branch, the longitudinal noise ${\cal N}_{\rm L}(z_0)/{\cal N}_0$ shows a quick saturation. For the highest frequency cut, we observe that the symmetric and antisymmetric modes are notably separated from each other. This results in the noise that seems to almost saturate just upon crossing the symmetric mode, which has notably larger momentum compared to the previous cut. In fact, the actual saturation happens much later, only after crossing the antisymmetric branch. Most remarkably, for the cut that passes just at the Josephson gap, the noise is significant at the smallest momenta due to the contribution from the antisymmetric mode and then gradually saturates on crossing the symmetric mode. This peculiar behavior is the signature of the gapped antisymmetric branch.

\section{Spin-structure of the superconducting pairing}
\label{sec:SpinN}

In this section, we turn to investigate spin fluctuations in the superconductor. Like current fluctuations, these also contribute to the magnetic noise, which according to Eq.~\eqref{eq:Tinv}, is related to the qubit depolarization rate. Except for a few rather exotic experimentally relevant systems, which we mention below, we find that the spin noise is often suppressed compared to the current noise. Importantly, by varying the orientation of the qubit quantization axis $\hat{\n}_q$, it is possible to sense the anisotropy of the spin noise, which can furnish useful information about the spin structure of the pairing wave function. Below we first discuss some broad qualitative features of the spin noise and then turn to the results of detailed microscopic calculations that justify our conclusions.

We start with crude estimates, similar to our discussion in Sec.~\ref{sec:Rel} of the current noise, and show that the spin noise is expected to be suppressed in metals compared to the current noise~\cite{Agarwal2017}. 
Spin fluctuations in the sample contribute to the local magnetic field $\B(\r_0)$ through the kernel $K_S(z_0) \sim 1/z_0^3$, arising from the long-range dipolar interaction between the impurity spin and the sample spins. We estimate the spin noise then to be:
\beq
\N_{\rm spin} & \approx & \mu_0^2 \int_{\cal{A}} d^2\r_1 \int_{\cal{A}} d^2\r_2K_{S}(\r_{01})K_{S}(\r_{02}) \nn
&&
 \qquad\qquad\qquad\qquad\qquad\qquad\quad
\times\langle \{  S(\r_1), S(\r_2)\} \rangle \nn 
& = & \frac{\mu_0^2}{z_0^{4}} \int d^2\r \langle \{S(\r),S(0)\}\rangle ,
\eeq
where we have used $|\r_{01}| \approx z_0 \approx |\r_{02}|$ and translational invariance of the spin-spin correlation function. Assuming the spin correlation length is smaller than the sample-probe distance $z_0$, one gets $\N_{\rm spin}\sim 1/z_0^4$. This suggests that the spin noise is suppressed by an extra factor of $1/z_0^2$ relative to the current noise (cf. Eq.~\eqref{eq:SchemNTL}).
For systems with a Fermi surface, the proper dimensionless ratio is $1/(k_F z_0)^2$, which is typically small.

In superconductors, a more careful analysis is required because of two observations. First, one might worry that our qualitative argument misses an enhancement of the spin noise for $T \lesssim T_c$ due to a conspiracy of coherence factors, which manifests as the Hebel-Slichter peak in NMR measurements~\cite{Tinkham,coleman_2015}. In the NMR case, one probes spatially local (or momentum integrated) spin-spin correlations, which effectively results in an energy integral over the square of the density of states~\cite{coleman_2015}. In case of clean s-wave superconductors, while $\nu(E) = E/\sqrt{E^2 - \Delta^2}$ has a weak singularity at $E = \Delta$, $\nu^2(E)$ has a much stronger non-integrable singularity, which is responsible for the development of the Hebel-Slichter peak. In contrast to the NMR probe, the impurity qubit is sensitive to non-local spin-spin correlations at momenta $q \sim 1/z_0$. Hence, one needs to integrate only over $\nu(E)$, which gives a non-singular result (up to logarithmic corrections, which are also present in $\N_{\rm T}$). 
Therefore, there is no anomalous enhancement of the spin noise.

Second, according to Eq.~\eqref{eq:Nz2}, at low temperatures the presence of a superflow suppresses the transverse current noise by an additional $1/z_0^2$ factor. We find that there is no analogous suppression of the spin noise, which could give a gateway for $\N_{\rm spin}$ to develop in the regime of strong superconductivity. If so, the anisotropy of the spin noise provides rich information about the pairing wave function. Still, as we demonstrate below in this subsection, the dimensionless ratio of the spin noise to the current noise is $(\mu_0 \mu_B \Lambda/e v_F)^2$, which is also typically quite small. It can become notable in superconductors emerging from flat bands with small Fermi velocities.

We now turn to a microscopic evaluation of the magnetic noise arising from spin fluctuations in a 2D material. Here we address only the question of clean superconductors and leave the disordered case for future work. \footnote{In contrast to the computation of the current response, the evaluation of the spin response for disordered superconductors requires taking into account disorder ladder diagrams.} The spin noise is related to the spin-spin correlation function as~\cite{CRD18,Joaquin_magnon_18} (we fix $\hat{\n}_q = \hat{z}$ and assume $\beta \Omega \ll 1$):
\beq
 \mathcal{N}_{\text{spin}}(\Omega) &=& \frac{(\mu_0 \mu_B)^2 k_B T}{128 \pi a^2 z_0^4 \Omega} \int_0^\infty dx \, x^3 e^{-x}  \times \bigg[C^{\prime \prime}_{zz}\left( \frac{x}{2z_0},\Omega \right)  \nn    
&+ &  \frac{1}{4} \bigg( C^{\prime \prime}_{-+}\left( \frac{x}{2z_0},\Omega \right) + C^{\prime \prime}_{+-}\left( \frac{x}{2z_0},\Omega \right) \bigg) \bigg].
\label{eq:NSpin}
\eeq
Here $a$ is the microscopic lattice spacing and $C^{\prime \prime}_{\alpha \beta}(\q, \Omega) \equiv - \text{Im}[ C_{\alpha \beta}(\q, \Omega)]$, where the retarded spin-spin correlator is defined as:
\begin{equation}
    {\cal C}_{\alpha\beta}(\r, t) = - i \Theta(t)
\langle [\sigma^\alpha(\r,t), \sigma^\beta(0,0)]\rangle
\end{equation}
and ${\cal C}_{\alpha\beta}(\q,\Omega) = \frac{1}{{\cal A}}\int_{-\infty}^{\infty} dt\, e^{i\Omega t} \int d^2\r \, e^{-i \q \cdot \r} {\cal C}_{\alpha\beta}(\r, t)$ represents the Fourier transform. Using Eq.~\eqref{eq:NSpin}, we now switch to evaluate the spin noise for singlet and triplet superconductors with different pairing wave functions. Below we focus on the main physical picture and relegate the computational details to Appendix~\ref{app:CSpin}.

\subsection{Singlet superconductors}

The spin noise in singlet superconductors reminds the transverse current noise, which we investigated in Sec.~\ref{sec:NT}. We note that due to the spin rotational symmetry,  $\mathcal{C}_{\alpha \beta} (\q, \Omega) = \delta_{\alpha \beta} \, \mathcal{C}(\q,\Omega)$. Hence, it is sufficient to compute only $\mathcal{C}_{zz} (\q, \Omega)$, which is proportional to the correlation function of $m_z(\q,\omega) = g \mu_B S_z(\q,\omega) \approx \mu_B \sigma_z(\q,\omega)$:
\beq
m_{z}(\q) &=& \mu_B \sum_{\k} ( c^{\dagger}_{\k - \q/2,\ua} c_{\k + \q/2, \ua} - c^{\dagger}_{\k - \q/2,\da} c_{\k + \q/2, \da} ) \nn 
&=& \mu_B \sum_{\k} \Psi^\dagger_{\k - \q/2} \Psi_{\k + \q/2}.
\eeq 
We note that the operator $m_z(\q)$ is similar to the current operator $j_\alpha(\q)$, with the only difference that the vertex factor $\mu_B$ is replaced by $e v_{\rm T}$. The correlator $\mathcal{C}(\q, \Omega)$ is then given by: 
\begin{align}
    \mathcal{C}^{\prime \prime}(\q, \Omega)  = \pi \Omega & \mu_B^2 a^2    \int \frac{d^2k}{(2\pi)^2} \int d\omega \; \left(-\frac{\partial n_F(\omega)}{\partial \omega}\right)\notag\\
& \quad \times \Tr[A(\k_-,\omega) A(\k_+, \omega + \Omega)]. \label{eq:CorrSpin}
\end{align}
Using the analysis of the transverse normal conductivity in Eq.~\eqref{eq:sigmaN}, we conclude that momentum and temperature scalings of $\mathcal{C}^{\prime \prime}(\q,\Omega)/\Omega$ are the same as for $\sigma_n^{\rm T}(\q,\Omega)$ (up to numerical prefactors).
In the regime of weak superconductivity, comparing Eqs.~\eqref{eq:Nz1} and \eqref{eq:sigmaN} with Eqs.~\eqref{eq:NSpin} and \eqref{eq:CorrSpin}, we deduce that the dimensionless ratio that determines the relative noise $\mathcal{N}_{\text{spin}}/\mathcal{N}_{\rm T}$ is $(\mu_B/e v_F z_0)^2 = \hbar^2/(2 m_e v_F z_0)^2 \sim 1/(k_F z_0)^2$, where we have assumed that $m_e \sim$ effective electron mass $m$. 
Typically $k_F z_0 \gg 1$, so that the spin noise is expected to be suppressed. 
For the regime of strong superconductivity, we obtain that the distance scalings of both types of noise are the same. To determine the ratio of their magnitudes, we now compare Eqs.~\eqref{eq:Nz2} and \eqref{eq:sigmaN} with Eqs.~\eqref{eq:NSpin} and \eqref{eq:CorrSpin} and find that $\mathcal{N}_{\text{spin}}/\mathcal{N}_{\rm T}\approx(\mu_0 \mu_B \Lambda/e v_F)^2 \sim (\mu_0 \mu_B n_s/m^* v_F)^2$.
Plugging in some typical values: $m^* = m_e$, $v_F = 10^5$\,m/s and $n_s = 10^{20}$\,m$^{-2}$, we get $\mathcal{N}_{\text{spin}}/\mathcal{N}_{\rm T} \sim 10^{-6}$.  Our analysis, therefore, shows that it is legitimate to disregard the spin noise compared to the one due to current fluctuations.
We remark that the spin noise might be notable for systems with small Fermi velocity or with additional contributions to superfluid stiffness arising from bands with significant Berry curvature~\cite{Torma}. 
In such flat-band systems with topological character~\cite{Andrei}, $\mathcal{N}_{\text{spin}}$ may be used to probe the spin structure of superconducting correlations.

\subsection{Triplet superconductors}

For triplet superconductors, the spin-spin correlation function becomes anisotropic and depends on the orientation of the order parameter vector $\bm{\Delta}_\k = \Delta \d_\k$ [see Appendix~\ref{app:CSpin} for additional discussion]:
\begin{widetext}
\beq
\mathcal{C}_{\alpha \beta}(\q, i\Omega_n) 
& = & \frac{1}{\beta V} \sum_{\k, i\omega_n} \Bigg\{ \frac{ \left[ 2 i \omega_n (i \omega_n + i \Omega_n) + 2\xi_{\k_+} \xi_{\k_-} +  \Delta^2(\d_{\k_-} \cdot \d^*_{\k_+} + \textrm{c.c.}) \right] \delta_{\alpha \beta}}{[(i \omega_n )^2 - E_{\k_-}^2][(i \omega_n + i \Omega_n)^2 - E_{\k_+}^2]} \nn
&& ~~~~~~~~~~~~~~~ + \frac{\Delta^2 [(d^\alpha_{\k_-} (d^{\beta}_{\k_+})^* + d^\alpha_{\k_+} (d^{\beta}_{\k_-})^* + \textrm{c.c.}) - 2 \delta_{\alpha \beta}(\d_{\k_-} \cdot \d^*_{\k_+} + \textrm{c.c.}) ]}{[(i \omega_n )^2 - E_{\k_-}^2][(i \omega_n + i \Omega_n)^2 - E_{\k_+}^2]} \Bigg\}.
\label{eq:TripletC}
\eeq
\end{widetext}
The first isotropic term in Eq.~\eqref{eq:TripletC} is analogous to the contribution from a singlet superconductor, while the second anisotropic term is special to a triplet superconductor and arises as a consequence of the broken SO(3) symmetry. By probing this term, for instance, by tuning the qubit quantization axis $\n_q$, one can distinguish different types of triplet superconductivity. Furthermore, one can distinguish triplet from singlet superconductors.
However, the anisotropic part of spin correlations does not contain any additional singularity in the coherence factors [see Appendix \ref{app:CSpin}]. Similar to the above discussion of singlet superconductors, the spin noise is expected to give a parametrically small contribution to the magnetic noise relative to the one from fluctuating currents, both in the weak and strong superconducting regimes.

\section{Conclusions and outlook}
\label{sec:conc}
We have discussed isolated impurity qubits, such as NV centers in diamond, as probes of superconductivity in two-dimensional materials. 
The qubit relaxation rate provides spatio-temporally resolved information about current correlations in the sample, the behavior of which sheds light on important intrinsic properties of the 2D sample of interest. We have shown the temperature dependence of the noise can signal the onset of superconductivity and can further be used to distinguish different superconducting gap structures. We have also demonstrated that the dependence of noise on the sample-probe distance probes different transport regimes in the superconducting state. By exploiting the suppression of transverse noise at low temperatures, we have shown how the qubit can be used to detect both spin noise and longitudinal collective modes. The former provides important additional information about the spin structure in the superconducting state, while the latter allows for the study of plasmon-like excitations. 

The qubit probe provides a novel route by which to investigate the rich physics associated with 2D superconductivity.  Various interesting fluctuation phenomena --  such as the interplay between superconducting fluctuations and disorder \cite{larkin2005theory,Stepanov2018}, superconducting phase fluctuations \cite{Emery1995}, Higgs modes \cite{Podolsky2011}, and Bardasis-Schrieffer modes\cite{Bardasis1961} -- could also conceivably be probed via qubit noise measurements, and the mean-field calculations presented here serve as a starting point for the more sophisticated analyses that would be required to model such effects. The local nature of the qubit probe may make it a useful tool in addressing questions of granular superconductivity, as is relevant, for instance, in the description of ``anomalous metals" \cite{Kivelson2019}.  Superconductivity under pressure \cite{Yanpeng2016,yankowitz2019} can also be readily probed via techniques described here, as certain qubits, such as NV centers, can be 
integrated into diamond anvil cells \cite{lesik2019,Satcher}.

\section*{Acknowledgements}
We thank T. Andersen, B. Dwyer, S. Hsieh, S. Kolkowitz, V. Manucharian, J. F. Rodriguez-Nieva, E. Urbach, R. Xue, A. Yacoby and C. Zu for helpful conversations.
S.C. was supported by the ARO through the Anyon Bridge MURI program (Grant No. W911NF-17-1-0323) via M. P. Zaletel, and the  U.S. DOE, Office of Science, Office of Advanced Scientific Computing Research, under the Accelerated Research in Quantum Computing (ARQC) program via N.Y. Yao. 
P.E.D, I.E., and E.D. were supported by Harvard-MIT CUA, AFOSR-MURI: Photonic Quantum Matter Award No. FA95501610323, Harvard Quantum Initiative, and AFOSR Grant No. FA9550-21-1-0216. N.Y.Y. acknowledges support from U.S. DOE, Office of Science, Office of Advanced Scientific Computing Research Quantum Testbed Program.
Support from the Gordon and Betty Moore Foundation is also gratefully acknowledged.

\bibliography{Sc_NV_LongPaper}

\onecolumngrid
\appendix
\section{Computation details for normal fluid conductivity}
\label{app:SigmaN}
In this appendix, we provide a detailed derivation of the normal-fluid transverse conductivity $\sigma_n^{\rm T}(\q, \Omega)$ for 2D superconductors, in clean and disordered limits. We remark that the screening effects due to the long-range Coulomb interaction affect the longitudinal normal conductivity, but not the transverse one (this can be easily seen within the two-fluid model in the main text). This justifies to compute the transverse conductivity within the BCS mean-field theory using the standard one-loop Kubo formula, which relates $\sigma_n^{\rm T}(\q, \Omega)$ to the transverse current-current correlation function  of the normal fluid:
\beq
\Pi_{\rm T}(\q, \tau) = - \frac{1}{V} \langle T_\tau (j_{\rm T}(\q, \tau) j_{\rm T}(-\q, 0)) \rangle_T,
\eeq
where $j_{\rm T}(\q, \tau) = (\hat{z} \times \hat{\q}) \cdot  \j^n(\q, \tau)$ is the transverse current, $V$ is the system volume (area in two dimensions), and $\langle \cdots \rangle_T$ denotes a thermal average in an equilibrium ensemble at temperature $T$. The real part of the conductivity can be obtained via analytic continuation of the Matsubara current-current correlation function as follows:
\beq
\text{Re}[\sigma_n^{\rm T} (\q, \Omega)] = - \frac{1}{\Omega} \text{Im}[\Pi_T(\q, \Omega)], \text{ where } \Pi_{\rm T}(\q, \Omega) \xrightarrow{i \Omega_n \rightarrow \Omega + i 0^+} \Pi_{\rm T}(\q, i \Omega_n) = \int_0^\beta d\tau \, e^{i \Omega_n \tau} \Pi_{\rm T}(\q, \tau).
\eeq

To capture the electromagnetic response of superconductors, we employ the two-fluid model. Within this framework, the normal fluid contribution to conductivity comes from quasiparticle excitations above the superconducting ground state, which we describe via the BCS mean-field theory\cite{Tinkham,coleman_2015}.
The mean-field BCS Hamiltonian of a singlet superconductor is given in terms of electron creation and annihilation operators $c^\dagger_{\k, \sigma}$ and $c_{\k, \sigma}$, dispersion $\xi_{\k} = \varepsilon_\k - \mu$, and gap-function $\Delta_\k$ as (we assume inversion or time-reversal symmetry, so that $\xi_{\k} = \xi_{-\k}$, and choose a gauge such that $\Delta_\k \in \mathbf{R}$):
\beq
H = \sum_{\k} \Psi^\dagger_\k h_\k \Psi_{\k}, \text{ where } \Psi_\k = \begin{pmatrix}
c_{\k, \uparrow} \\
c^\dagger_{-\k, \downarrow} \end{pmatrix} \text{ and } h_\k = \begin{pmatrix} \xi_\k & \Delta_\k \\ \Delta_\k & -\xi_{\k} \end{pmatrix}.
\eeq
Denoting the quasiparticle excitation energy by $E_\k = \sqrt{\xi_\k^2 + \Delta_\k^2}$ and introducing a phenomenological lifetime via a self-energy $\Sigma(\k,i\omega_n)$, one obtains the Matsubara Green's function ($\omega_n = (2n+1)\pi/\beta$):
\beq
G(\k, i \omega_n) &=& (i \omega_n - \Sigma_{\k,i\omega_n} - h_\k)^{-1} \nn &=& \frac{1}{(i\omega_n - \Sigma(\k,i\omega_n))^2 - E^2_\k} \begin{pmatrix} i \omega_n - \Sigma(\k,i\omega_n) + \xi_\k & \Delta_\k \\ \Delta_\k & i \omega_n - \Sigma(\k,i\omega_n) - \xi_{\k} \end{pmatrix}.
\label{eq:Gfswave}
\eeq
The evaluation of the pair correlator is conveniently carried out via the spectral function representation of the Green functions~\cite{altland2010condensed}:
\beq
G(\k,i \omega_n) = \int_{-\infty}^\infty d\omega' \frac{A(\k, \omega')}{i \omega_n - \omega'}, \; \; A(\k, \omega) = - \frac{1}{\pi} \text{Im}[G^R(\k,\omega)], 
\eeq
where  $G^R(\k, \omega)$ is the retarded Green function (it can be obtained from the Matsubara one by proper analytical continuation~\cite{altland2010condensed}). Within a simple model of isotropic disorder scattering, the self-energy $\Sigma^R(\k, \omega)$ can be approximated as $\Sigma^R(\k, \omega \rightarrow 0) \approx -i \Gamma_0$, where $\Gamma_0$ is the isotropic scattering rate of electrons at the Fermi surface (we assume that the real part of $\Sigma^{R}(\k,\omega)$ just renormalizes the bare dispersion). In this limit, we have:
\beq
\label{eq:generalSF}
A(\k, \omega) = -\frac{(\omega^2 - \Gamma_0^2 - E_{\k}^2) \Gamma_0}{\pi[(\omega^2 - \Gamma_0^2 - E_{\k}^2)^2 + (2\omega \Gamma_0)^2]} \begin{pmatrix} 1 & 0 \\ 0 & 1 \end{pmatrix} + \frac{2 \omega \Gamma_0}{\pi[(\omega^2 - \Gamma_0^2 - E_{\k}^2)^2 + (2\omega \Gamma_0)^2]} \begin{pmatrix} \omega + \xi_\k & \Delta_{\k} \\ \Delta_{\k} & \omega - \xi_\k \end{pmatrix} \label{eq:Akw}.
\eeq

To evaluate the dissipative part of the conductivity $\text{Re}[\sigma_{\alpha \beta} (\q, \Omega)]$ within linear response, we need to consider only the paramagnetic part of the current operator, which is given in terms of the spinor $\Psi_\k$ by 
\beq
j_{\alpha}(\q) = e \sum_{\k} v_\alpha(\k) c^{\dagger}_{\k - \q/2,\sigma} c_{\k + \q/2, \sigma} = e \sum_{\k} v_\alpha(\k) \Psi^\dagger_{\k - \q/2} \Psi_{\k + \q/2}, \text{ where } \bm v(\k) = \frac{\partial \varepsilon_\k}{\partial \k}. 
\eeq
Using $\k_{\pm} = \k \pm \q/2$, we find:
\beq
\Pi_{\alpha \beta}(\q, i\Omega_n) &=&  \frac{e^2}{\beta V}\sum_{\k, i \omega_n} v_{\alpha} v_{\beta}\Tr[G(\k_-,i\omega_n) G(\k_+,i\omega_n + i \Omega_n) ] \nn
& = & \frac{e^2}{V}\sum_{\k} \int d\omega_1 \int d\omega_2 \; v_{\alpha} v_{\beta} \Tr[A(\k_-,\omega_1) A(\k_+, \omega_2)] \left( \frac{n_F(\omega_1) - n_F(\omega_2)}{i \Omega_n + \omega_1 - \omega_2} \right). 
\eeq
Accordingly, the conductivity is given by (taking the continuum limit):
\beq \sigma_{\alpha \beta}(\q,\Omega) &=&  -\frac{\text{Im}[\Pi_{\alpha \beta}(\q, i\Omega_n \rightarrow \Omega + i 0^+)]}{\Omega} \nn &=& \frac{e^2 \pi}{\Omega} \int \frac{d^2k}{(2\pi)^2} \int d\omega_1 \int d\omega_2 \; v_{\alpha} v_{\beta} \Tr[A(\k_-,\omega_1) A(\k_+, \omega_2)] \left(n_F(\omega_1) - n_F(\omega_2) \right)\delta(\Omega + \omega_1 - \omega_2)  \label{eq:sigma1} \\
 &=&  e^2 \pi  \int \frac{d^2k}{(2\pi)^2} \int d\omega \; v_{\alpha} v_{\beta} \Tr[A(\k_-,\omega) A(\k_+, \omega + \Omega)] \left( \frac{n_F(\omega) - n_F(\omega + \Omega)}{\Omega}  \right). \label{eq:sigma2}
\eeq
Let us introduce $\hat{\bm{t}} = \hat{z} \times \hat{\q}$. Then the transverse conductivity is defined as $\sigma^{\rm T}(\q,\Omega) = \hat{t}_\alpha \sigma_{\alpha\beta}(\q,\Omega)\hat{t}_\beta$. 
In the clean limit, it is more convenient to use Eq.~\eqref{eq:sigma1} to carry out the computation, while in the dirty limit, Eq.~\eqref{eq:sigma2} is more convenient. 

\emph{Simplifications in the clean limit:} 
The expression for conductivity can be further simplified in the clean limit ($\Gamma_0 \to 0$), where the first term in Eq.~\eqref{eq:Akw} vanishes, while the second term reduces to a sum of delta functions:
\beq
A(\k,\omega) = \frac{1}{2 E_\k} \left( \delta(\omega - E_\k) - \delta(\omega + E_\k) \right) \begin{pmatrix} \omega + \xi_\k & \Delta_\k \\ \Delta_\k & \omega - \xi_\k \end{pmatrix}
\label{eq:cleanSF}.
\eeq
Plugging this into Eq.~\eqref{eq:sigma1} and evaluating the trace and the integrals over $\omega_1$ and $\omega_2$, we get:
\beq
\sigma_{\alpha \beta}(\q, \Omega) & = &  \frac{e^2 \pi}{2 \Omega} \int \frac{d^2k}{(2\pi)^2} v_\alpha v_\beta \bigg[ \left(1 + \frac{\xi_+ \xi_- + \Delta_+ \Delta_-}{E_+ E_-} \right) \left( n_F(E_+) - n_F(E_-) \right) \left[  \delta(\Omega + E_+ - E_-) - \delta(\Omega - E_+ + E_-) \right]  \nn
& + & \left(1 - \frac{\xi_+ \xi_- + \Delta_+ \Delta_-}{E_+ E_-} \right) (1 - n_F(E_+) - n_F(E_-)) \left[ \delta(\Omega - E_+ - E_-) - \delta(\Omega + E_+ + E_-) \right].
\label{eq:sigma3}
\eeq
where by $\pm$ we have indicated momenta $\k \pm \q/2$. 
For the physically reasonable case of small $q \ll k_F$, we can further approximate $\xi_+ \xi_- + \Delta_+ \Delta_- \approx E_+ E_- + O(q^2)$, so the terms in the second line of Eq.~(\ref{eq:sigma3}) can be neglected. The energy constraint further simplifies to 
\beq
E_- - E_+ = -\left[ \q \cdot \bm v(\k) \left( \frac{\xi_\k}{E_\k} \right) + \q \cdot \bm v_\Delta(\k) \left( \frac{\Delta_\k}{E_\k} \right) \right],
\eeq
where $\bm v_\Delta(\k) = \partial_\k \Delta_\k$ is the gap-velocity. 
Further assuming inversion symmetry, we can reduce Eq.~(\ref{eq:sigma3}) to the following form:
\beq
\sigma_{\alpha \beta}(\q, \Omega) = \frac{e^2}{2\pi} \int d^2k \; v_\alpha v_\beta  \frac{n_F(E_\k) - n_F(E_\k + \Omega)}{\Omega}  \delta\left(\Omega + \q \cdot \bm v(\k) \left( \frac{\xi_\k}{E_\k} \right) + \q \cdot \bm v_\Delta(\k) \left( \frac{\Delta_\k}{E_\k} \right) \right).
\label{eq:sigma4}
\eeq
Equation~\eqref{eq:sigma4} is the most general expression for the conductivity tensor in the clean limit, which we will subsequently use to evaluate the transverse conductivity for different types of superconductors.

\emph{Simplifications in the dirty limit:} In the dirty limit ($\Gamma_0 \gg k_B T \gtrsim \omega$), the spectral functions are smooth on the scale of $\omega$ due to disorder-smearing. Hence, in the physically relevant regime of $\beta \Omega \ll 1$, we can make the following approximation:
\beq
\frac{n_F(E_\k) - n_F(E_\k + \Omega)}{\Omega} \approx - \frac{\partial n_F}{\partial \omega} \approx \delta(\omega).
\eeq
In this limit, the expression for conductivity in Eq.~\eqref{eq:sigma2} reduces to:
\beq
\sigma_{\alpha \beta}(\q, \Omega) = e^2 \pi  \int \frac{d^2k}{(2\pi)^2} v_\alpha v_\beta \Tr[A(\k_-,0) A(\k_+, 0)]. 
\eeq
We note from Eq.~\eqref{eq:Akw} that in the $\omega \to 0$ limit, the second term vanishes, and the first term acquires a particularly simple Lorentzian form:
\beq
A(\k,0) = \frac{\Gamma_0}{\pi(\Gamma_0^2 + E_\k^2)} \begin{pmatrix} 1 & 0 \\ 0 & 1 \end{pmatrix}.
\eeq
Accordingly, the simplified expression we will use for the conductivity tensor in the disordered limit is:
\beq
\sigma_{\alpha \beta}(\q, \Omega) = \frac{e^2}{2\pi}  \int d^2k \; v_\alpha v_\beta \frac{\Gamma_0}{\pi(\Gamma_0^2 + E_+^2)} \frac{\Gamma_0}{\pi(\Gamma_0^2 + E_-^2)}.
\label{eq:sigma8}
\eeq

\subsection{s-wave superconductors}
We now focus on s-wave superconductors with $\Delta_\k = \Delta$ and zero gap-velocity. 
For the transverse conductivity, we need the component of velocity  perpendicular to $\q$.
If we assume a circular Fermi surface (arising from a quadratic dispersion $\xi_\k = \frac{k^2}{2m} - \mu$) with $\bm v_F = v_F \hat{k}$ and take $\theta$ as the angle between vectors $\k$ and $\q$, we get $v_{\rm T}(\hat{k}) \approx v_F \sin\theta$. 
Thus, in the clean limit, we can simplify Eq.~\eqref{eq:sigma4} as follows:
\beq
\sigma^{\rm T}(\q, \Omega) &=& \frac{e^2 v_F^2}{2 \pi} \int_0^\infty dk \, k \left( \frac{n_F(E_\k) - n_F(E_\k + \Omega)}{\Omega} \right) \int_0^{2\pi}  d\theta \, \sin^2\theta \;  \delta \left(\Omega + \frac{q v_F \xi_\k}{E_\k}\cos \theta \right) \nn
&=& \frac{e^2 v_F^2 \beta}{\pi q v_F} \int_0^\infty dk \,  \frac{k E_\k}{4|\xi_\k| \cosh^2( \beta E_\k/2)} \sqrt{ 1 - \left( \frac{\Omega E_\k}{q v_F \xi_\k}\right)^2} \, \Theta\left( \frac{|\xi_\k| v_F q}{E_\k} - \Omega \right).
\label{eq:sigma5}
\eeq
Approximating $\xi_{\k} \approx v_F(k - k_F)$, we see that the angular integral over $\theta$ requires $v_F q > \Omega$ and places further restrictions on allowed momenta k for a given $\Omega$:
\beq
\Omega \leq \frac{q v_F |\xi_\k|}{E_\k} = \frac{q v_F^2 |k - k_F|}{\sqrt{v_F^2 |k - k_F|^2 + \Delta^2}} \implies v_F|k-k_F| \geq \frac{\Delta \Omega}{\sqrt{v_F^2 q^2 - \Omega^2}}.
\eeq
It is convenient to introduce a dimensionless variable $\alpha \equiv \Omega/q v_F$, which is typically small $\alpha \sim \Omega d/v_F \ll 1$. We can also scale all dimensionful quantities out of the $k$-integral using $y \equiv (k - k_F)\xi_T$, where $\xi_T = v_F/\Delta(T)$ is the superconducting coherence length.
The lower limit for the $y$ integral is $-k_F \xi_T$, which can be approximated by $-\infty$ as the coherence length is typically much larger than $k_F^{-1}$. 
Further, at low temperatures, much smaller than the Fermi energy, the derivative of the Fermi function constrains us to remain close to the Fermi surface. 
Therefore, we can replace the factor of $k$ in the numerator of Eq.~\eqref{eq:sigma5} by $k_F$, and we find the following approximate expression for the transverse conductivity
\beq
\sigma^{\rm T}(\q, \Omega) \approx \frac{e^2 v_F k_F \beta \Theta(1 - \alpha)}{\pi q \xi_{T}} \int_{|y| \geq \frac{\alpha}{\sqrt{1 - \alpha^2}}} dy \, \frac{\sqrt{(1 + y^2)(1 - \alpha^2(1+y^2)/y^2)}}{4 |y| \cosh^2(\beta \Delta \sqrt{1+ y^2}/2)} 
\label{eq:sigma6}.
\eeq
We carefully note that as $\alpha \to 0$ (corresponding to $\Omega \to 0$), there is a weak logarithmic divergence of $\sigma^{\rm T}$ in $\alpha$. 
This arises from the singular density of states at the gap threshold, and we expect such a divergence to be cured by the presence of a small disorder. 
At low temperatures, the integral scales as $e^{-\beta \Delta(T)}$, as can be seen by approximating the $\cosh^2$ in the denominator by an exponential:
\beq
\sigma^{\rm T}(\q, \Omega) \approx \frac{e^2 v_F k_F \beta e^{-\beta \Delta } \Theta(1 - \alpha)}{\pi q \xi_{T}} \int_{|y| \geq \frac{\alpha}{\sqrt{1 - \alpha^2}}} dy \, \frac{\sqrt{(1 + y^2)(1 - \alpha^2(1+y^2)/y^2)}}{|y| } e^{-\beta \Delta y^2/2}.
\eeq
We also note that as we increase temperature $T$ to approach $T_c$, the gap goes to zero and $\xi_T$ diverges. In this limit, the integral should not be non-dimensionalized with $y = (k - k_F)\xi_T$, but rather just evaluated from Eq.~\eqref{eq:sigma5} directly in the metallic limit where $\xi_\k = E_\k$. Doing so leads to the following expression for the transverse conductivity of the normal metal:
\beq
\sigma^{\rm T}(\q, \Omega) \approx \frac{e^2 k_F \sqrt{1 - \alpha^2} \, \Theta(1 - \alpha)}{\pi q} = \frac{e^2 v_F k_F \beta \sqrt{1 - \alpha^2}\, \Theta(1 - \alpha)}{\pi q \lambda_T} .
\label{eq:sigma7}
\eeq
The static ($\Omega = q v_F \alpha \to 0$) limits of Eqs.~\eqref{eq:sigma6} and \eqref{eq:sigma7} appear in Eqs.~\eqref{eq:SigmaSwave} and \eqref{eq:SigmaCleanMetal} respectively in the main text.

In the dirty limit, we employ Eq.~\eqref{eq:sigma8} to evaluate the transverse conductivity (defining $\gamma^2 \equiv \Gamma_0^2 + \Delta^2$ to simplify notations):
\beq
&& \sigma^{\rm T}(\q, \Omega \to 0) = \frac{e^2}{2\pi^3}  \int d^2k \; v^2_{\rm T} \left(\frac{\Gamma_0}{\Gamma_0^2 + E_+^2} \right)\left(\frac{\Gamma_0}{\Gamma_0^2 + E_-^2} \right) \nn 
&& \approx \frac{e^2 v_F^2 \Gamma_0^2}{2\pi^3} \int_0^\infty dk \, k \, \int_0^{2\pi}d\theta \, \frac{\sin^2\theta}{\left( \gamma^2 +  \xi_\k^2 + \xi_\k v_F q \cos\theta + (v_F q/2)^2 \right)\left( \gamma^2 + \xi_\k^2 - \xi_\k v_F q \cos\theta + (v_F q/2)^2 \right)} \nn
&& = \frac{e^2 v_F^2 \Gamma_0^2}{4\pi^2} \int_0^\infty dk \, k \, \frac{\gamma_0^2 + \xi_\k^2 + (v_F q/2)^2 - \sqrt{(\xi_\k^2 - (v_F q/2)^2)^2 + 2(\xi_\k^2 + (v_F q/2)^2)\gamma^2 + \gamma^4}}{\xi_\k^2(v_F q/2)^2 (\gamma^2 + \xi_\k^2 + (v_F q/2)^2)}  \nn
&& = \frac{e^2 v_F^2 \Gamma_0^2}{\pi^2} \int_0^\infty dk \, k \, \frac{1}{(\gamma^2 + \xi_\k^2 + (v_F q/2)^2)(\gamma^2 + \xi_\k^2 + (v_F q/2)^2 + \sqrt{(\xi_\k^2 - (v_F q/2)^2)^2 + 2(\xi_\k^2 + (v_F q/2)^2)\gamma^2 + \gamma^4})} .\nn \label{eqn:sigma_dirty}
\eeq
While the above expression is quite opaque, substantial simplifications can be made in two limits. The first limit is $\gamma^2 = \Gamma_0^2 + \Delta^2 \ll (v_F q)^2$, which corresponds to $z_0 \ll \ell_{\rm MF}, \xi_T$ or a small sample-probe distance (but still much larger than the average electron separation scale $\bar{a}$ so that $q/k_F \sim \bar{a}/z_0 \ll 1$). In this limit, the conductivity can be evaluated by setting $\gamma^2 = \Gamma_0^2 + \Delta^2 = 0$ in Eq.~\eqref{eqn:sigma_dirty} (we also set $\xi = v_F(k - k_F)$ and $k = k_F$ in the numerator):
\beq
\sigma^{\rm T}(\q, \Omega \to 0) &\approx& \frac{e^2 v_F k_F \Gamma_0^2}{\pi^2} \int_{-k_F v_F}^\infty d\xi \, \frac{1}{( \xi^2 +  (v_Fq/2)^2 + |\xi^2 -  (v_Fq/2)^2| ) \left(\xi^2 + (v_Fq/2)^2 \right)} \nn
& \approx & \frac{e^2 v_F k_F \Gamma_0^2}{\pi^2} \int_{-\infty}^\infty d\xi \, \frac{1}{2 \, \text{max}\{ \xi^2, (v_Fq/2)^2  \} \left(\xi^2 + (v_Fq/2)^2 \right)} \nn
& = & \frac{e^2 v_F k_F \Gamma_0^2}{\pi^2} \left[ \frac{1}{(v_Fq/2)^2} \int_0^{v_Fq/2} d\xi \, \frac{1}{\xi^2 + (v_Fq/2)^2} + \int_{v_Fq/2}^{\infty} d\xi \, \frac{1}{\xi^2 (\xi^2 + (v_Fq/2)^2)}  \right] \nn
& = & \frac{e^2 v_F k_F \Gamma_0^2}{\pi^2 (v_Fq/2)^3}.
\label{eq:sigma11}
\eeq
Thus, we have analytically derived the $1/q^3$ scaling of $\sigma^{\rm T}(\q,0)$ that we expected from arguments in the main text. The other limit where we can evaluate expression~\eqref{eqn:sigma_dirty} is the large sample-probe distance scenario with $z_0 \gg \ell_{\rm MF}$. In this case, we can neglect all $q$ dependence in the denominator and we find that:
\beq
\sigma^{\rm T}(\q, \Omega \to 0) & = & \frac{e^2 v_F^2}{2 \pi^3} \int_0^\infty dk \, k \, \frac{\Gamma_0^2}{(\Gamma_0  +\Delta^2 + \xi_\k^2)^2} \int_0^{2\pi} d\theta \, \sin^2\theta 
 \approx  \frac{e^2 v_F k_F \Gamma_0^2}{2 \pi^2} \int_{-\infty}^{\infty}d\xi \, \frac{1}{(\Gamma_0^2  +\Delta^2 + \xi^2)^2} \nn
& = & \frac{e^2 v_F k_F \Gamma_0^2}{4\pi (\Gamma_0^2  +\Delta^2)^{3/2}}.
\label{eq:sigma12}
\eeq
which is Eq.~\eqref{eq:sigmaSDis} in the main text. 

A simple interpolating function $\tilde{\sigma}$ which captures both these limits may be guessed by looking at the expressions for $\sigma^{\rm T}(\q,0)$ in Eqs.~\eqref{eq:sigma11} and \eqref{eq:sigma12}:
\beq
\tilde{\sigma}(\q) = \frac{e^2 v_F k_F \Gamma_0^2}{4\pi} \frac{1}{\left[ \Gamma_0^2 + \Delta^2 + \left( \frac{\pi}{4} \right)^{2/3} \left( \frac{v_F q}{2} \right)^{2} \right]^{3/2}}.
\eeq
In practice, $\tilde{\sigma}(q)$ approximates $\sigma^{\rm T}(q,\Omega \to 0)$ remarkably well for a large range of parameter values (we checked this numerically).
Since a naive substitution of $\sigma^{\rm T}(q \sim 1/z_0)$ in Eq.~\eqref{eq:Nz1} gives an unphysical result that the $\mathcal{N}_{\rm T}$ increases with increasing $z_0$ (a consequence of the integral being highly singular as $q \to 0$), we instead use $\tilde{\sigma}(q)$ to deduce the behavior of $\mathcal{N}_{\rm T}$ for small $z_0$. 
We find that in the weak superconducting regime, $\mathcal{N}_{\rm T} \sim \text{const.} - z_0 \ln(z_0\sqrt{\Gamma_0^2 + \Delta^2}/v_F)$ for small $z_0 \ll \ell_{\rm MF}, \xi_T$, justifying the distance dependence discussed in the main text.

\subsection{d-wave superconductors}
Let us now consider a d-wave superconductor, with $\Delta_\k = \Delta(T) (\cos k_x - \cos k_y)$. The quasiparticle dispersion then has nodal points along $k_x = \pm k_y$, and at each nodal point, the Fermi velocity $\bm v_F = \nabla_\k \xi_\k$ (or the normal to the Fermi surface) is perpendicular to the gap velocity $ \bm v_\Delta = \nabla_\k \Delta_\k$. Therefore, at each node $\k_0$, $\bm  v_F$ and $ \bm v_\Delta$ form a local orthogonal basis, and we can write the quasiparticle energy as $E_\k = \sqrt{v_F^2 k_\parallel^2 + v_\Delta^2 k_\perp^2}$, where $\k - \k_0 = (k_\parallel,k_\perp)$. 
To evaluate the conductivity, it is convenient to scale out the anisotropy of the Dirac cones by rescaling the momenta at each node, as was done in Ref.~\onlinecite{DurstLee}. Defining vectors $\tilde{\k} \equiv (v_F k_\parallel,v_\Delta k_\perp)$ and $\tilde{\q} \equiv (v_F q_\parallel,v_\Delta q_\perp)$, the integral measure around $\k_0$ is given by:
\beq
\int d^2k \to \int dk_\parallel dk_\perp \to \int \frac{d^2\tilde{k}}{v_F v_\Delta}.
\eeq
We first consider the clean limit, given by Eq.~\eqref{eq:sigma4}. Energy conservation takes a convenient form in the rescaled coordinates.
\beq
\Omega = E_- - E_+ = -\left[ \q \cdot \v_F \left( \frac{\xi_\k}{E_\k} \right) + \q \cdot \v_\Delta \left( \frac{\Delta_\k}{E_\k} \right) \right] = -\tilde{q} \cos(\theta_{\tilde{\q}} - \theta_{\tilde{\k}}).
\eeq
One might worry that rescaling momenta in this way results in added complications for the transverse component of velocity $v_\alpha(\hat{k})$.
A nice simplification occurs in the angular integral by noting that  $v_\alpha(\hat{k})$ is entirely set by the angle between $\q$ and $\k_0$ at the node $\k_0$. 
Since $\k_0$ rotates by $\pi/2$ as we cycle through the 4 nodes, for fixed $\q$ the angle $\theta_{\k_0,\q}$ also increases by $\pi/2$ and therefore we have:
\beq
\sum_{\rm nodes} v_{\rm T}^2(\hat{k}) = 2 v_F^2 (\sin^2 \theta_{\k_0,\q} + \cos^2\theta_{\k_0,\q}) = 2 v_F^2.
\eeq
Therefore, after summing the contributions from all four nodes, the conductivity in the clean limit is given by (noting that $\tilde{k} = E_\k$ and $\tilde{q} = E_\q$):
\beq
\sigma^{\rm T}(\q,\Omega) &=&  \frac{e^2 v_F}{\pi v_\Delta} \int_0^\infty d\tilde{k} \, \tilde{k} \frac{\beta e^{\beta \tilde{k}}}{(e^{\beta \tilde{k}} + 1)^2} \int_0^{2\pi} d\theta_{\tilde{k}} \delta(\Omega + \tilde{q} \cos(\theta_{\tilde{\q}} - \theta_{\tilde{\k}}) ) \nn
& = & \frac{2e^2 v_F}{\pi v_\Delta} \int_0^\infty d\tilde{k} \, \tilde{k} \frac{\beta e^{\beta \tilde{k}}}{(e^{\beta \tilde{k}} + 1)^2} \frac{\Theta(\tilde{q} - \Omega)}{\sqrt{ \tilde{q}^2 - \Omega^2 }}
 = \frac{2e^2 v_F \ln(2)}{\beta \pi v_\Delta} \frac{\Theta(\tilde{q} - \Omega)}{\sqrt{ \tilde{q}^2 - \Omega^2 }} \xrightarrow{\Omega \to 0} \frac{2e^2 v_F \ln(2)}{\beta \pi v_\Delta E_\q}. 
 \label{eq:sigma9}
\eeq
Therefore, the transverse noise in this limit is given by (approximating $\coth(x) = 1/x$ for small $x$):
\beq
\mathcal{N}_{\rm T}(\Omega \to 0) &=& \frac{2\mu_0^2 e^2 v_F \ln(2)}{2 \pi \beta^2 v_\Delta} \int \frac{d^2 \tilde{q}}{(2\pi)^2 v_F v_\Delta} \frac{ e^{- 2 |\q| z_0} }{\tilde{q}} \nn 
&=& \frac{2\mu_0^2 e^2 T^2 \ln(2)}{(2 \pi)^3 v_\Delta^2} \int_0^{2\pi} d\theta_{\tilde{q}} \int_{0}^\infty d \tilde{q} \, \exp\left[- 2 \tilde{q} z_0 \sqrt{\left(\frac{\sin \theta}{v_\Delta}\right)^2 + \left(\frac{\cos \theta}{v_F} \right)^2 }\right] \nn
&=& \frac{2\mu_0^2 e^2 T^2 \ln(2)}{(2 \pi)^3 z_0 v_\Delta} \left[ K\left( 1- \frac{v_\Delta^2}{v_F^2} \right) +\frac{v_F}{v_\Delta} K\left( 1- \frac{v_F^2}{v_\Delta^2} \right)  \right] \xrightarrow{v_\Delta \ll v_F} \frac{4 \mu_0^2 e^2 T^2 \ln(2)}{(2 \pi)^3 z_0 v_\Delta} \ln\left( \frac{4 v_F}{v_\Delta} \right),  \label{eq:sigma10}
\eeq
where $K(x) = \int_0^{\pi/2} d\theta (1 - x \sin^2\theta)^{-1/2}$ is the elliptic integral. Equations~\eqref{eq:sigma9} and \eqref{eq:sigma10} correspond to Eqs.~\eqref{eq:SigmaDwaveClean} and \eqref{eq:NoiseDwaveClean} in the main text.

In the dirty limit, we once again resort to Eq.~\eqref{eq:sigma8} to evaluate the transverse conductivity. Just like in the clean case, we can sum over the transverse velocity squared over all nodes:
\beq
\sigma^{\rm T} (\q,\Omega \to 0) &\approx& \frac{e^2 v_F^2 \Gamma_0^2}{ \pi^3} \int_0^\infty \frac{d\tilde{k} \; \tilde{k}}{v_F v_\Delta} \int_0^{2 \pi} d\theta \frac{1}{(\Gamma_0^2 + \tilde{k}^2 + \tilde{q}^2/4)^2 - \tilde{k}^2 \tilde{q}^2 \cos^2(\theta)} \nn
& = & \frac{2 e^2 v_F \Gamma_0^2}{v_\Delta \pi^2}   \int_0^\infty d\tilde{k} \, \frac{ \tilde{k} }{(\Gamma_0^2 + \tilde{k}^2 + \tilde{q}^2/4)\sqrt{(\Gamma_0^2 + \tilde{k}^2 + \tilde{q}^2/4)^2 - \tilde{k}^2 \tilde{q}^2}} \nn
& = & \frac{2e^2 v_F}{\pi^2 v_\Delta} \frac{\Gamma_0^2 \ln\left( 1 + \frac{\tilde{q}(\tilde{q} + \sqrt{4\Gamma_0^2 + \tilde{q}^2})}{2 \Gamma_0^2} \right)}{ \tilde{q}\sqrt{4\Gamma_0^2 + \tilde{q}^2} }  = \frac{e^2 v_F}{\pi^2 v_\Delta} \frac{4 \Gamma_0^2 \, \sinh^{-1}\left( \frac{\tilde{q}}{2\Gamma_0} \right)}{ \tilde{q}\sqrt{4\Gamma_0^2 + \tilde{q}^2}}.
\eeq 
For $\tilde{q} = (q_\parallel^2 v_F^2 + q_\perp^2 v_\Delta^2)^{1/2} \rightarrow 0$ limit, we recover the Durst-Lee result of a constant universal conductivity $e^2 v_F/\pi^2 v_\Delta$, independent of the disorder strength~\cite{DurstLee}.
In the opposite limit for large $\tilde{q}/\Gamma_0$, we have:
\beq
\sigma^{\rm T} (\q,\Omega \to 0) \approx \frac{e^2 v_F}{\pi^2 v_\Delta} \frac{4 \ln(\tilde{q}/\Gamma_0)}{(\tilde{q}/\Gamma_0)^2}.
\eeq
This $1/q^2$ scaling of conductivity leads to a sample-probe distance independent noise (up to logarithmic corrections) in the weak superconductor. 
To show this, we carry an approximate integral over momenta $\q$ (using Eqs.~\eqref{eq:NT} and \eqref{eq:rs} in the main text), with a lower cutoff on $q$ set by $\Gamma_0/\sqrt{v_F^2 + v_\Delta^2}$:
\beq
\mathcal{N}_{\rm T}(\Omega) \sim \Gamma_0^2 \; {\rm Ei} \left(\frac{4 \Gamma_0 z_0}{\sqrt{v_F^2 + v_\Delta^2}} \right),
\eeq
where $\text{Ei}(x) = \int_{x}^{\infty} dt \; \frac{e^{-t}}{t}$.
For small $x$, $\text{Ei}(x) \sim \ln(x)$, indicating that the transverse noise in the weak superconducting phase is independent of $z_0$ up to logarithmic corrections, as discussed in the main text.  

\subsection{Triplet superconductors}

In this section, we turn to fill in a few missing details regarding the transverse conductivity calculation in triplet superconductors. 
We recall that the BCS Hamiltonian for a p-wave triplet superconductor~\cite{coleman_2015} can be written in terms of the four-component BW spinor $\Psi_\k = (c_\k,\, i\sigma^y c^\dagger_{-\k})^T = (c_{\k,\ua},\, c_{\k,\da},\, c_{-\k,\da}^\dagger,\, -c_{-\k,\ua}^\dagger)^T$ as:
\beq
H_{\rm BCS} = \sum_{\k \in \frac{1}{2} BZ} \Psi^\dagger_\k h_\k \Psi_\k,\quad h_\k &=& \xi_\k \tau^z + (\bm \Delta_\k \cdot \bm{\sigma}) \tau^+ + (\bm \Delta_\k^* \cdot \bm{\sigma}) \tau^-. 
\eeq 
Here $\tau^{\pm} = \frac{1}{2}(\tau^x \pm i \tau^y)$ act in the Nambu (particle-hole) space. $\bm{\Delta}_\k = \Delta \, \d_\k$ represents the superconducting order parameter, with $\d_\k$ being appropriately normalized over the Fermi surface.
Assuming unitary pairing with $\d_{\k} \parallel \d^*_{\k}$, the Matsubara Green's function is given by:
\beq
G(\k, i \omega_n) = (i \omega_n - \Sigma_{\k,i\omega_n} - h_\k)^{-1}  = \frac{i \omega_n - \Sigma_{\k,i\omega_n} + \xi_\k \tau^z + (\bm \Delta_\k \cdot \bm\sigma) \tau^+ + (\bm \Delta_\k^* \cdot \bm \sigma) \tau^-}{(i \omega_n - \Sigma_{\k,i\omega_n})^2 - E_\k^2} 
\label{eq:Gfpwave},
\eeq
where $E_\k = \sqrt{\xi_\k^2 + \Delta^2 |\bm d_\k|^2}$. Likewise in the discussion above, we again assume an isotropic disorder-induced scattering rate of electrons at low-energy, i.e, $\Sigma_{\k,\omega} \equiv \Sigma(\k, i\omega_n \rightarrow \omega + i 0^+) = - i \Gamma_0$. To evaluate the normal-fluid transverse conductivity $\sigma^T(\q, \Omega)$, we employ Eqs.~\eqref{eq:sigma1} and~\eqref{eq:sigma2}.

In the clean limit, $\Gamma_0 \to 0$, $\Pi_{\alpha \beta}(\q, i \omega_n)$ is given by (dividing by a factor of 2 so that we can extend the summation over $\k$ to the entire BZ):
\beq
\Pi_{\alpha \beta}(\q, i \Omega_n) &=&  \frac{e^2}{2\beta V} \sum_{\k \in BZ, i\omega_n} v_\alpha v_\beta \Tr[ G(\k_-, i \omega_n) G(\k_+, i \omega_n + i \Omega_n)] \nn
& = & \frac{e^2}{\beta V} \sum_{\k, i\omega_n} v_\alpha v_\beta \frac{2 i \omega_n (i \omega_n + i \Omega_n) + 2\xi_{\k_+} \xi_{\k_-} +  \Delta^2(\d_{\k_-} \cdot \d^*_{\k_+} + \textrm{c.c.})}{[(i \omega_n )^2 - E_{\k_-}^2][(i \omega_n + i \Omega_n)^2 - E_{\k_+}^2]}.\label{eqn::triplet_clean_PI}
\eeq
The key difference between a singlet and triplet superconductor is that the latter has a nontrivial spin structure of the pairing function, as exemplified by $(\d_{\k_-} \cdot \d^*_{\k_+} + \textrm{c.c.})$ in Eq.~\eqref{eqn::triplet_clean_PI}. We consider two different kinds of p-wave pairings:
$\d_\k =  (k_x \hat{x} + k_y \hat{y})/k_F $ and $\d_\k = \hat{z}(k_x + i k_y)/k_F$. 
In both cases, we note that near the Fermi surface, where $k \approx k_F$ and $q \ll k_F$, we have:
\beq
\d_{\k_-} \cdot \d^*_{\k_+} + \textrm{c.c.} = \frac{2}{k_F^2}\left(\k^2 - \frac{\q^2}{4}\right) \approx 2.
\label{eq:dmdotdp}
\eeq
Therefore, in both cases, we have:
\beq
\Pi_{\alpha \beta}(\q, i \Omega_n) = \frac{2e^2}{\beta V} \sum_{\k, i\omega_n} v_\alpha v_\beta \frac{i \omega_n (i \omega_n + i \Omega_n) + \xi_{\k_+} \xi_{\k_-} +  \Delta^2}{[(i \omega_n )^2 - E_{\k_-}^2][(i \omega_n + i \Omega_n)^2 - E_{\k_+}^2]}.
\eeq
Carrying out the Matsubara summation~\cite{altland2010condensed}, we find that:
\beq
\Pi_{\alpha \beta}(\q, i \Omega_n) & = & \frac{e^2}{2V}\sum_{\k} v_\alpha v_\beta \bigg[ \left( 1 + \frac{\xi_+ \xi_- + \Delta^2}{E_+ E_-} \right) (n_F(E_+) - n_F(E_-)) \left(\frac{1}{i \Omega_n - E_- + E_+} - \frac{1}{i \Omega_n + E_- - E_+} \right) \nn
& &+ \left( 1 - \frac{\xi_+ \xi_- + \Delta^2}{E_+ E_-} \right) (1 - n_F(E_-) - n_F(E_+)) \left( \frac{1}{i \Omega_n - E_- - E_+} - \frac{1}{i\Omega_n + E_- + E_+} \right) \bigg].
\eeq
Performing the analytic continuation to the real axis $i\Omega_n \to \Omega + i0^{+}$ and taking the imaginary part of $\Pi_{\alpha \beta}$ , we find that the expression for the real part of the conductivity tensor reduces to that of a fully-gapped isotropic s-wave superconductor, which has already been computed above, cf. Eq.~\eqref{eq:sigma3}.

A similar conclusion also holds for the case with a finite disorder $\Gamma_0 \neq 0$, where the alternate representation~\eqref{eq:sigma2} is more convenient to evaluate the resulting Matsubara sums.

\section{Computation details for the  spin-spin correlation function}
\label{app:CSpin}
In this section, we provide detailed derivation of the magnetic noise for singlet and triplet superconductors. 
We begin by recalling the definition of the noise tensor:
\beq
\N_{ab}(\Omega) = \frac{1}{2} \int_{-\infty}^{\infty} dt \ e^{i \Omega t} \langle \{ B_a(\r_0,t) , B_b(\r_0,0) \} \rangle,
\eeq
where $B_a(\r_0,t)$ is the local, mediated by spin fluctuations in the sample, magnetic field at the location of the qubit. For concreteness, we further assume that the qubit quantization axis $\hat{\bm{n}}_q \parallel \hat{z}$; generalization to an arbitrary direction of $\hat{\bm{n}}_q$ is straightforward. 
The qubit depolarization rate is then set by $\N_{-+}(\Omega)$, which is related to spin correlations in the sample as~\cite{CRD18,Joaquin_magnon_18}:
\beq
\N_{\rm spin}(\Omega) \xrightarrow{\hat{\bm{n}}_q = \hat{z}} \N_{-+}(\Omega) = \frac{(\mu_0 \mu_B)^2}{16 \pi a^2} \coth\left(\frac{\beta \Omega}{2} \right) \int_0^\infty dq \, q^3 e^{-2 q z_0} \left[ \frac{1}{4}\left( C^{\prime \prime}_{-+}(q,\Omega) + C^{\prime \prime}_{+-}(q,\Omega) \right) + C^{\prime \prime}_{zz}(q,\Omega) \right],
\label{eq:NspinApp}
\eeq
where $C^{\prime \prime}_{\alpha \beta}(\q, \Omega) \equiv - \text{Im}[ C_{\alpha \beta}(\q, \Omega)]$, the latter being the equilibrium retarded spin-spin correlation functions at temperature $T$: 
\beq
{\cal C}_{\alpha\beta}(\q,\Omega) = -\frac{i}{V}\int_{0}^{\infty} dt\, e^{i\Omega t} \int d^2\r \, e^{-i \q \cdot \r}\langle [\sigma^\alpha(\r,t), \sigma^\beta(0,0)]\rangle_T,
\eeq
where $\sigma^\alpha$ are the Pauli matrices (we have assumed $S = 1/2$). 
The retarded spin-spin correlation function can be obtained by analytic continuation of the corresponding Matsubara correlation function:
\beq
{\cal C}_{\alpha\beta}(\q,i\Omega_n) \xrightarrow{i \Omega_n \to \Omega + i0^+} {\cal C}_{\alpha\beta}(\q,\Omega). 
\eeq
Therefore, the task at hand is to compute ${\cal C}_{\alpha\beta}(\q,i\Omega_n)$ for microscopic models of singlet and triplet superconductors. 

\subsection{Singlet superconductors}
For singlet superconductors, SO(3) spin-rotational symmetry implies that ${\cal C}_{\alpha\beta}(\q,i\Omega_n) = \delta_{\alpha \beta} {\cal C}(\q,i\Omega_n)$.
Therefore, it is sufficient to calculate ${\cal C}_{zz}(\q,i\Omega_n)$, which is given in terms of the Nambu spinors, cf. Appendix \ref{app:SigmaN}, as:
\beq
\sigma^z(\q, i \omega_n) = \sum_{\k,\sigma = \pm} \sigma c^\dagger_{\k - \q/2, \sigma} c_{\k + \q/2, \sigma} =   \sum_{\k} \Psi^\dagger_{\k -\q/2} \Psi_{\k + \q/2}.
\eeq
Using the Matsubara Green's function $G(\k,i \omega_n)$ for s-wave superconductors defined in Eq.~\eqref{eq:Gfswave}, the spin-spin correlator can be expressed as:
\begin{gather}
    {\cal C}(\q,i\Omega_n) = {\cal C}_{zz}(\q,i\Omega_n)  = \frac{a^2}{\beta V} \sum_{\k, i\omega_n} \Tr[G(\k_-,i\omega_n) G(\k_+,i\omega_n + i \Omega_n) ] \nn
\implies \frac{{\cal C}^{\prime \prime}(\q,\Omega)}{\Omega}  =  \frac{\pi a^2}{V}\sum_{\k} \int d\omega \; \Tr[A(\k_-,\omega) A(\k_+, \omega + \Omega)] \left( -\frac{\partial n_F}{\partial \omega} \right).
\end{gather}

Using ${\cal C}_{+-} + {\cal C}_{-+} = 2 ( {\cal C}_{xx} + {\cal C}_{yy}) = 2 {\cal C}$ in Eq.~\eqref{eq:NspinApp}, we find that the spin noise is given by:
\beq
\N_{\rm spin}(\Omega) &=& \frac{(\mu_0 \mu_B)^2}{16 \pi a^2} \coth\left(\frac{\beta \Omega}{2} \right) \int_0^\infty dq \, q^3 e^{-2 q z_0} [ 2 {\cal C}^{\prime \prime}(q,\Omega)] \nn
& \xrightarrow{\beta \Omega \ll 1} & \frac{(\mu_0 \mu_B)^2}{4 \pi \beta a^2} \int_0^\infty dq \, q^3 e^{-2 q z_0} \left[ \frac{{\cal C}^{\prime \prime}(q,\Omega)}{\Omega} \right] \nn
& = & \frac{(\mu_0 \mu_B)^2}{4 \beta} \int_0^\infty dq \, q^3 e^{-2 q z_0} \int \frac{d^2k}{(2\pi)^2} \int d\omega \; \Tr[A(\k_-,\omega) A(\k_+, \omega + \Omega)] \left( -\frac{\partial n_F}{\partial \omega} \right)
\label{eq:Nspin2app}.
\eeq

We recall that in the weak superconducting regime, the noise due to transverse current fluctuations is given by:
\beq
\N_{\rm T}(\Omega) = \frac{\mu_0^2 e^2}{4 \beta} \int_0^\infty dq \, q e^{- 2 q z_0} \int \frac{d^2k}{(2\pi)^2}  v_{\rm T}^2 \int d\omega \; \Tr[A(\k_-,\omega) A(\k_+, \omega + \Omega)] \left( -\frac{\partial n_F}{\partial \omega} \right)
\label{eq:NTapp}.
\eeq
We note that if we approximate $v_{\rm T} \approx v_F$ in Eq.~\eqref{eq:NTapp}, then barring constant factors, the $\k$ and $\omega$ integrals in Eqs.~\eqref{eq:Nspin2app} and \eqref{eq:NTapp} start to look identical, while the $q$ integral sets the momentum scale to be $\sim 1/z_0$. Therefore, distance and temperature scalings of the spin noise for singlet superconductors can be inferred directly from $\N_{\rm T}$ discussed earlier, at least in the clean limit (where one is not worried about the disorder ladder diagrams contributing to the spin noise).
In particular, the temperature dependence of the spin noise in singlets is identical to $\N_{\rm T}$, while the distance scaling can be obtained by multiplying $\N_{\rm T}(z_0)$ by a factor of $1/z_0^2$.
To compare the magnitudes of the noises from spin and transverse current fluctuations, we scale all momenta by $z_0$ so that the momentum integrals become dimensionless. Approximating $v_{\rm T} \approx v_F$ and $m_e \approx$ effective electron mass $m$, we find:
\beq
\frac{\N_{\rm spin}}{\N_{\rm T}} \simeq \left( \frac{\mu_B}{e v_F z_0} \right)^2 = \left( \frac{\hbar}{2 m_e v_F z_0} \right)^2 \approx \left( \frac{1}{2k_F z_0} \right)^2.
\eeq

In the regime of strong superconductivity, as a consequence of the noise suppression by the superflow, cf. Eqs.~\eqref{eq:Nz1} and \eqref{eq:Nz2}, $\N_{\rm T}$ contains an additional factor of $1/(\mu_0 z_0 \Lambda)^2$ compared to the case of weak superconductors.
Therefore, in this regime, the ratio is given by:
\beq
\frac{\N_{\rm spin}}{\N_{\rm T}} \simeq \left( \frac{\mu_0 \mu_B \Lambda}{e v_F} \right)^2 .
\eeq
While $\N_{\rm spin}$ and $\N_{\rm T}$ carry the same distance dependence in this regime, the overall dimensionless ratio is still quite small (see the main text for estimates).

\subsection{Triplet superconductors}
For triplet superconductors, the spin-spin correlation function is anisotropic and depends on the orientation of $\bm{\Delta}_\k$ that transforms as a vector under SO(3) spin rotations. 
In the clean limit, the correlator $\mathcal{C}_{\alpha \beta}(\q, \Omega)$ can be computed using the Green's function from Eq.~\eqref{eq:Gfpwave}:
\beq
\mathcal{C}_{\alpha \beta}(\q, i\Omega_n)  &=& \frac{1}{2\beta V} \sum_{\k \in BZ, i\omega_n} \Tr[ \sigma^\alpha G(\k_-, i \omega_n) \sigma^\beta G(\k_+, i \omega_n + i \Omega_n)] \nn
& = & \frac{1}{\beta V} \sum_{\k, i\omega_n} \Bigg\{ \frac{ \left[ 2 i \omega_n (i \omega_n + i \Omega_n) + 2\xi_{\k_+} \xi_{\k_-} +  \Delta^2(\d_{\k_-} \cdot \d^*_{\k_+} + \textrm{c.c.}) \right] \delta_{\alpha \beta}}{[(i \omega_n )^2 - E_{\k_-}^2][(i \omega_n + i \Omega_n)^2 - E_{\k_+}^2]} \nn
&& ~~~~~~~~~~~~~~~ + \frac{\Delta^2 [(d^\alpha_{\k_-} (d^{\beta}_{\k_+})^* + d^\alpha_{\k_+} (d^{\beta}_{\k_-})^* + \textrm{c.c.}) - 2 \delta_{\alpha \beta}(\d_{\k_-} \cdot \d^*_{\k_+} + \textrm{c.c.}) ]}{[(i \omega_n )^2 - E_{\k_-}^2][(i \omega_n + i \Omega_n)^2 - E_{\k_+}^2]} \Bigg\},
\label{eq:TripletCapp_app}
\eeq
where we have used the Pauli matrix identity: $\Tr[\sigma^a\sigma^b\sigma^c\sigma^d] = 2(\delta_{ab}\delta_{cd} - \delta_{ac}\delta_{bd} + \delta_{ad}\delta_{bc})$. 
The first term in Eq.~(\ref{eq:TripletCapp_app}) is analogous to the contribution from a singlet superconductor, while the second term is special to a triplet superconductor --- its real part to be interpreted as the polarizability of the condensate.
Using $\d_{\k_-} \cdot \d^*_{\k_+} + \textrm{c.c.} \approx 2$ for both kinds of triplet pairing we consider (see Eq.~\eqref{eq:dmdotdp}), we can further simplify the expression in Eq.~\eqref{eq:TripletCapp_app} to:
\beq
\mathcal{C}_{\alpha \beta}(\q, i\Omega_n) = \frac{1}{\beta V} \sum_{\k, i\omega_n} \frac{ \left[ 2 i \omega_n (i \omega_n + i \Omega_n) + 2\xi_{\k_+} \xi_{\k_-} + 2 \Delta^2 \right] \delta_{\alpha \beta} + \Delta^2 [d^\alpha_{\k_-} (d^{\beta}_{\k_+})^* + d^\alpha_{\k_+} (d^{\beta}_{\k_-})^* + \textrm{c.c.} - 4 \delta_{\alpha \beta} ]}{[(i \omega_n )^2 - E_{\k_-}^2][(i \omega_n + i \Omega_n)^2 - E_{\k_+}^2]}.\label{eq:C_generic_1}  ~~~~~~~
\eeq

\emph{ BW case:} We first consider the BW pairing wave function with $\bm{d}_\k = (k_x \hat{x} + k_y \hat{y})/k_F$ and quasiparticle energy $E_{\k} = \sqrt{\xi_\k^2 + \Delta^2}$.
In this case, we have $d^\alpha_{\k_-} (d^{\beta}_{\k_+})^* + \textrm{c.c.} \approx 2 \hat{k}_\alpha \hat{k}_\beta \approx d^\alpha_{\k_+} (d^{\beta}_{\k_-})^* + \textrm{c.c.}.$, for $k \approx k_F$ and $q \ll k_F$. Here $\alpha, \beta = x, y$ (otherwise, we will get zero if either of $\alpha$ or $\beta$ is equal to $z$). From this, we compute the imaginary part of the retarded spin-spin correlation function, obtained from Eq.~\eqref{eq:C_generic_1} by analytic continuation $i \Omega_n \to \Omega + i0^+$:
\begin{align}
\mathcal{C}^{\prime \prime}_{\alpha \beta}(\q, \Omega) &  = \frac{\pi}{2} \int \frac{d^2k}{(2\pi)^2} \Bigg\{ \left[ \left( 1 + \frac{\xi_+ \xi_- +  \Delta^2 }{E_+ E_-} \right) \delta_{\alpha \beta} + \frac{2\Delta^2 (\hat{k}_\alpha \hat{k}_\beta - \delta_{\alpha \beta})}{E_+ E_-} \right] [n_F(E_+)- n_F(E_-)] \notag\\
& \times (\delta(\Omega - E_- + E_+) - \delta(\Omega + E_- - E_+)) + \left[ \left( 1 - \frac{\xi_+ \xi_- +  \Delta^2 }{E_+ E_-} \right) \delta_{\alpha \beta} - \frac{2\Delta^2 (\hat{k}_\alpha \hat{k}_\beta - \delta_{\alpha \beta})}{E_+ E_-} \right]  \notag\\
&\times [1 - n_F(E_+) - n_F(E_-)][\delta(\Omega - E_- - E_+) - \delta(\Omega + E_- + E_+)]\Bigg\}.
\end{align}
To evaluate the angular integral, it is convenient to choose an orthogonal set of in-plane directions with $\alpha = \beta = \hat{q}$ (we will call this correlator $\mathcal{C}_{q}$) and with $\alpha = \beta = \hat{z} \times \hat{q}$ (we will call this correlator $\mathcal{C}_{t}$). 
Since we only need $\mathcal{C}^{\prime \prime}_{q}+ \mathcal{C}^{\prime \prime}_{t}$ to evaluate $\N_{\rm spin}$ (this is because we chose $\hat{\bm{n}}_q = \hat{z}$), we find that:
\beq
\mathcal{C}^{\prime \prime}_{q}(\q, \Omega) + \mathcal{C}^{\prime \prime}_{t}(\q, \Omega) = \pi \int \frac{d^2k}{(2\pi)^2} \left( 1 + \frac{\xi_+ \xi_- }{E_+ E_-} \right) (n_F(E_+)- n_F(E_-))(\delta(\Omega - E_- + E_+) - \delta(\Omega + E_- - E_+)) \nn
+ \left( 1 - \frac{\xi_+ \xi_-}{E_+ E_-} \right) (1 - n_F(E_+) - n_F(E_-))(\delta(\Omega - E_- - E_+) - \delta(\Omega + E_- + E_+)).~~~~~~~~
\eeq
We now specialize on positive frequencies $\Omega > 0$, so that the last delta function $\delta(\Omega + E_- + E_+)$ does not contribute. 
The delta function $\delta(\Omega - E_- - E_+)$ only contributes when $\Omega \geq 2 \Delta$, i.e, when the probe frequency is larger than twice the quasiparticle gap. Therefore this term can be neglected in the limit $\Omega \to 0$ (we remark that even though there is an inverse square-root singularity at $\Omega = 2 \Delta$, this regime is not expected to show up in experiments since $\Omega$ is typically much smaller than all other energy scales). 
Further, using the inversion symmetry and neglecting terms of $O(q^2)$ in the numerator, we can reduce the integral to a form similar to what we have already calculated for the transverse conductivity:
\begin{align}
\frac{\mathcal{C}^{\prime \prime}_{q}(\q, \Omega) + \mathcal{C}^{\prime \prime}_{t}(\q, \Omega)}{\Omega} &= \frac{1}{2 \pi}  \int_0^\infty dk \, k \, \left(1 +  \frac{\xi_\k^2}{E_\k^2} \right)  \frac{n_F(E_\k) - n_F(E_\k + \Omega)}{\Omega} \int_0^{2\pi}  d\theta \,  \delta \left(\Omega + \frac{q v_F \xi_\k}{E_\k}\cos \theta \right) \nn
&\approx \frac{\beta k_F}{\pi q v_F} \int_0^\infty dk \, \left(1 +  \frac{\xi_\k^2}{E_\k^2} \right)  \frac{E_\k}{4|\xi_\k| \cosh^2( \beta E_\k/2)} \frac{1}{\sqrt{ 1 - \left( \frac{\Omega E_\k}{q v_F |\xi_\k|}\right)^2}} \, \Theta\left( \frac{|\xi_\k| v_F q}{E_\k} - \Omega \right) \nn
& \approx \frac{\beta k_F}{\pi q v_F^2} \Theta(1 - \alpha) \int_{|\xi| \geq \frac{ \Delta \alpha}{\sqrt{1 - \alpha^2}}} d\xi \,   \frac{2\xi^2 + \Delta^2}{\xi^2 + \Delta^2}  \frac{\sqrt{\xi^2 + \Delta^2}}{4 |\xi| \cosh^2(\beta\sqrt{\xi^2 + \Delta^2}/2)}  \frac{1}{\sqrt{1 - \alpha^2(\xi^2 + \Delta^2)/\xi^2}} \nn
& =  \frac{\beta k_F}{\pi \xi_T q v_F} \Theta(1 - \alpha) \int_{|y| \geq \frac{\alpha}{\sqrt{1-\alpha^2}}} dy \, \frac{2 y^2 + 1}{|y| \sqrt{(1+y^2)(1 - \alpha^2(1 + y^{-2}))}} \frac{1}{4 \cosh^2(\beta \Delta \sqrt{1+y^2}/2)},
\end{align}
where we have defined $y = \xi/\Delta = (k - k_F)\xi_T$. We note that there is a weak logarithmic singularity in the correlation function as $\alpha = \Omega/(q v_F) \to 0$; it scales as $1/q$ (up to non-essential logarithms) and decays exponentially on decreasing $T$. 
Thus, the essential aspects of the behavior of $(\mathcal{C}^{\prime \prime}_q + \mathcal{C}^{\prime \prime}_{t})/\Omega$ are similar to the transverse normal conductivity of clean s-wave superconductors. 
Next, we consider $\mathcal{C}_{zz}(\q,\Omega)$: 
\beq
\frac{\mathcal{C}^{\prime \prime}_{zz}(\q,\Omega)}{\Omega} &=& \frac{1}{4\pi} \int_0^\infty dk \, k \, \left(1 +  \frac{\xi_\k^2 - \Delta^2}{E_\k^2} \right)  \frac{n_F(E_\k) - n_F(E_\k + \Omega)}{\Omega}  \int_0^{2\pi}  d\theta \,  \delta \left(\Omega + \frac{q v_F \xi_\k}{E_\k}\cos \theta \right) \nn
& \approx & \frac{\beta k_F}{\pi q v_F^2} \Theta(1 - \alpha) \int_{|\xi| \geq \frac{ \Delta \alpha}{\sqrt{1 - \alpha^2}}} d\xi \,  \frac{|\xi|}{\sqrt{\xi^2 + \Delta^2}}  \frac{1}{4 \cosh^2(\beta\sqrt{\xi^2 + \Delta^2}/2)}  \frac{1}{\sqrt{1 - \alpha^2(\xi^2 + \Delta^2)/\xi^2}} \nn
& = & \frac{\beta k_F}{\pi \xi_T q v_F} \Theta(1 - \alpha) \int_{|y| \geq \frac{\alpha}{\sqrt{1-\alpha^2}}} dy \, \frac{|y|}{\sqrt{(1 + y^2)(1 - \alpha^2(1 + y^{-2}))}} \frac{1}{4 \cosh^2(\beta \Delta \sqrt{1+y^2}/2)}. \label{eqn:app_Czz_BW}
\eeq
The $\mathcal{C}_{zz}$ correlator still scales as $1/q$ and decreases exponentially upon decreasing $T$, but there is no logarithmic singularity as $\alpha \to 0$. 
Thus, the spin noise will be highly anisotropic at low frequencies.
Combining the results for the in-plane and out-of-plane spin correlators, cf. Eq.~\eqref{eq:NspinApp}, one can deduce the spin noise for the BW triplet superconductor.

\emph{ ABM case:} Next, we consider the ABM pairing wave function with $\d_\k = \hat{z}(k_x + i k_y)/k_F$ and quasiparticle energy $E_\k = \sqrt{\xi_\k^2 + \Delta^2}$. 
For this pairing function, we have $d^\alpha_{\k_-} (d^\beta_{\k_+})^* + \textrm{c.c.}. \approx 2 \approx d^\alpha_{\k_+} (d^\beta_{\k_-})^* + \textrm{c.c}.$, for $k \approx k_F$ and $q \ll k_F$. Here $\alpha = \beta = z$, and we get zero otherwise. Following the same steps as above, we arrive at:
\begin{align}
    \mathcal{C}^{\prime \prime}_{\alpha \beta}(\q, \Omega) &= \frac{\pi}{2} \int \frac{d^2k}{(2\pi)^2} \Bigg\{ \left[ \left( 1 + \frac{\xi_+ \xi_- + \Delta^2}{E_+ E_-} \right) \delta_{\alpha \beta} + \frac{2 \Delta^2}{E_+ E_-} \left( \delta_{\alpha,z} \delta_{\beta,z} - \delta_{\alpha \beta} \right) \right](n_F(E_+)- n_F(E_-)) \nn 
& \times (\delta(\Omega - E_- + E_+) - \delta(\Omega + E_- - E_+)) + \left[ \left( 1 - \frac{\xi_+ \xi_- + \Delta^2}{E_+ E_-} \right) \delta_{\alpha \beta} - \frac{2 \Delta^2}{E_+ E_-} \left( \delta_{\alpha,z} \delta_{\beta,z} - \delta_{\alpha \beta} \right) \right] \nn
& \times(1 - n_F(E_+)- n_F(E_-))(\delta(\Omega - E_- - E_+) - \delta(\Omega + E_- + E_+))\Bigg\}.
\end{align}
For $\alpha = \beta = x$ or $y$, neglecting terms of $O(q^2)$ and assuming $\Omega < 2 \Delta$, we find that:
\beq
\frac{\mathcal{C}^{\prime \prime}_{xx}(\q,\Omega)}{\Omega} &=& \frac{\mathcal{C}^{\prime \prime}_{yy}(\q,\Omega)}{\Omega} = \frac{1}{4\pi} \int_0^\infty dk \, k \, \left(1 +  \frac{\xi_\k^2 - \Delta^2}{E_\k^2} \right)  \frac{n_F(E_\k) - n_F(E_\k + \Omega)}{\Omega} \int_0^{2\pi}  d\theta \,  \delta \left(\Omega + \frac{q v_F \xi_\k}{E_\k}\cos \theta \right) \nn
& = & \frac{\beta k_F}{\pi \xi_T q v_F} \Theta(1 - \alpha) \int_{|y| \geq \frac{\alpha}{\sqrt{1-\alpha^2}}} dy \, \frac{|y|}{\sqrt{(1 + y^2)[]1 - \alpha^2(1 + y^{-2})]}} \frac{1}{4 \cosh^2(\beta \Delta \sqrt{1+y^2}/2)},
\eeq
which is exactly identical to the expression for $\mathcal{C}^{\prime \prime}_{zz}(\q,\Omega)/\Omega$ for  the BW pairing, cf. Eq.~\eqref{eqn:app_Czz_BW}, and, thus, shares all its features. 
Finally, for $\alpha = \beta = z$, we obtain:
\beq
\frac{\mathcal{C}^{\prime \prime}_{zz}(\q,\Omega)}{\Omega} &=& \frac{1}{4\pi} \int_0^\infty dk \, k \, \left(1 +  \frac{\xi_\k^2 + \Delta^2}{E_\k^2} \right)  \frac{n_F(E_\k) - n_F(E_\k + \Omega)}{\Omega}  \int_0^{2\pi}  d\theta \,  \delta \left(\Omega + \frac{q v_F \xi_\k}{E_\k}\cos \theta \right) \nn
&=& \frac{1}{2\pi} \int_0^\infty dk \, k \,  \frac{n_F(E_\k) - n_F(E_\k + \Omega)}{\Omega}  \int_0^{2\pi}  d\theta \,  \delta \left(\Omega + \frac{q v_F \xi_\k}{E_\k}\cos \theta \right) \nn
& = & \frac{\beta k_F}{\pi \xi_T q v_F} \Theta(1 - \alpha) \int_{|y| \geq \frac{\alpha}{\sqrt{1-\alpha^2}}} dy \, \frac{\sqrt{(1 + y^2)}}{|y|\sqrt{1 - \alpha^2(1 + y^{-2})}} \frac{1}{4 \cosh^2(\beta \Delta \sqrt{1+y^2}/2)}.
\eeq
The same expression holds for singlet superconductors, which we discussed in the previous subsection. 
In order to get the full spin noise $\N_{\rm spin}$, one needs to sum all the contributions in accordance with Eq.~\eqref{eq:NspinApp}. 
Although the spin noise shares a lot in common with the transverse current noise, $\N_{\rm spin}$ turns out to be suppressed for both kinds of triplet pairings we consider (for further discussion, we refer to the main text).

\section{Longitudinal collective modes in bilayers}
\label{app:CM_bilayer}

\subsection{Collective modes}

In this appendix, we provide computational details of the longitudinal collective modes in bilayer superconductors and discuss their implications for noise. We decompose vectors as $\bm A(\bm q, z) = A_\parallel \hat{\bm q} + A_\perp \hat{\bm q} \times \hat{\bm z} + A_z \hat{\bm z}$. 
Since longitudinal modes couple with p-polarized waves, it 
allows us to set
$A_\perp = 0$ (we note that $\E = -\partial_t \A$ lies in the plane perpendicular to $\hat{z} \times \hat{q}$).
Therefore, the solutions (evanescent waves) to Maxwell's equations are given by:
\begin{align}
    & A_\parallel (\bm q, z;\omega) = \begin{cases} 
    {\cal A}_{1,\parallel} e^{-\varkappa (z - l/2)}, & \frac{l}{2} < z\\
    \alpha e^{-\varkappa_l z} + \beta e^{\varkappa_l z} , & -\frac{l}{2} < z < \frac{l}{2}\\
    {\cal A}_{2\parallel} e^{\varkappa (z + l/2)}, &  z < -\frac{l}{2}
    \end{cases},
\end{align}
where $\varkappa = \sqrt{q^2 -\epsilon\omega^2/c^2}$ and $\varkappa_l = \sqrt{q^2 -\epsilon_l\omega^2/c^2}$. ${\cal A}_{1,\parallel}(\bm q,\omega)$ and ${\cal A}_{2,\parallel}(\bm q,\omega)$ are the longitudinal components of the in-plane vector potentials. At this stage, $\alpha(\bm q,\omega)$ and  $\beta(\bm q,\omega)$ are yet unknown coefficients. 
Since there is no free charge density outside of the two layers, we have:
\beq
\nabla \cdot \A = i q A_{\parallel} + \partial_z A_z = \frac{1}{i\omega} \nabla \cdot \E = 0, ~~ z \neq \pm l/2. ~~~
\eeq
Using this and the Maxwell equation relating $\A$ to the interlayer current density $J_z$, we obtain:
\begin{align}
     A_z  = \begin{cases}  \frac{iq}{\varkappa}
    {\cal A}_{1,\parallel} e^{-\varkappa (z - l/2)}, &  \frac{l}{2} < z\\
    \frac{iq }{\varkappa_l} (\alpha e^{-\varkappa_l z} - \beta e^{\varkappa_l z}) + \frac{\mu_0}{\varkappa_l^2} J_z , & -\frac{l}{2} < z < \frac{l}{2}\\
    -\frac{iq}{\varkappa}{\cal A}_{2,\parallel} e^{\varkappa (z +l/2)}, &  z < -\frac{l}{2}
    \end{cases}.
\end{align}
Next, we use the continuity of the tangential component of the vector potential at the two-boundaries to relate the coefficients $\alpha$ and $\beta$ to ${\cal A}_{1,\parallel}$ and ${\cal A}_{1,\parallel}$:
\begin{align}
   & \alpha  = \frac{ {\cal A}_{2,\parallel} e^{\varkappa_l l/2} - {\cal A}_{1,\parallel} e^{-\varkappa_l l/2} }{e^{\varkappa_l l} - e^{-\varkappa_l l} }, & \beta  = \frac{ {\cal A}_{1,\parallel} e^{\varkappa_l l/2} - {\cal A}_{2,\parallel} e^{-\varkappa_l l/2} }{e^{\varkappa_l l} - e^{-\varkappa_l l} }.
\end{align}
From this result, we can relate $\vartheta$ to the in-plane vector potentials and the interlayer current density:
\begin{align}
    \vartheta = \frac{2\pi}{\Phi_0} \int_{-l/2}^{l/2} dz \, A_z =  \frac{2 \pi \mu_0 l }{\Phi_0 \varkappa_l^2}J_z + \frac{2\pi iq}{\Phi_0 \varkappa_l^2} ({\cal A}_{2,\parallel} - {\cal A}_{1,\parallel}).
    \label{eq:vartheta_bl}
\end{align}
Finally, we use the integral forms of Maxwell's equations across the two layers to relate the two-dimensional charge and current densities to the in-plane vector potentials and the interlayer current density:
\begin{align}
    \rho_1 = \epsilon_0 \omega q \Big[ \frac{\epsilon_l}{\varkappa_l}\frac{{\cal A}_{2,\parallel} - {\cal A}_{1,\parallel}\cosh \varkappa_l l }{\sinh \varkappa_l l} - \frac{\epsilon}{\varkappa} {\cal A}_{1,\parallel} \Big]  - \frac{i\omega \epsilon_l }{c^2\varkappa_l^2}J_z,\, \\
    \rho_2  = \epsilon_0 \omega q \Big[ \frac{\epsilon_l}{\varkappa_l}\frac{{\cal A}_{1,\parallel} - {\cal A}_{2,\parallel}\cosh \varkappa_l l }{\sinh \varkappa_l l} - \frac{\epsilon}{\varkappa} {\cal A}_{2,\parallel} \Big]  + \frac{i\omega \epsilon_l }{c^2\varkappa_l^2}J_z,\\
     j_{1,\rm L}  = \epsilon_0 \omega^2 \Big[ \frac{\epsilon_l}{\varkappa_l}\frac{{\cal A}_{2,\parallel} - {\cal A}_{1,\parallel}\cosh \varkappa_l l }{\sinh \varkappa_l l} - \frac{\epsilon}{\varkappa} {\cal A}_{1,\parallel} \Big]  - \frac{i q}{\varkappa_l^2}J_z,\, \\
     j_{2,\rm L}  = \epsilon_0 \omega^2 \Big[ \frac{\epsilon_l}{\varkappa_l}\frac{{\cal A}_{1,\parallel} - {\cal A}_{2,\parallel}\cosh \varkappa_l l }{\sinh \varkappa_l l} - \frac{\epsilon}{\varkappa} {\cal A}_{2,\parallel} \Big]  + \frac{i q}{\varkappa_l^2}J_z.
     \label{eq:densities_bl}
\end{align}
These results are consistent with the continuity equations~\eqref{eqn:cnt_1} and~\eqref{eqn:cnt_2}. 
Using the above equations on 10 variables $\{ \rho_{i}, j_{i,\rm L},  {\cal A}_{i,\parallel}, \theta_{i}, \vartheta, J_z\}$, one may compute the spectrum of the longitudinal collective modes, as we discuss below. 

\subsubsection{Symmetric mode}

We consider the symmetric combinations: $\theta_s = \theta_1 + \theta_2$, $\rho_s = \rho_1 + \rho_2$, $j_s = j_{1,\parallel} + j_{2,\parallel}$, ${\cal A}_s = {\cal A}_{1,\parallel} + {\cal A}_{1,\parallel}$. We then obtain a closed set of equations solely on these new symmetric variables:
\begin{align*}
    & \rho_s  = \epsilon_0 \omega q \Big[ \frac{\epsilon_l}{\varkappa_l}\frac{1 - \cosh \varkappa_l l }{\sinh \varkappa_l l} - \frac{\epsilon}{\varkappa} \Big]{\cal A}_s,\,  & j_s  = \epsilon_0 \omega^2 \Big[ \frac{\epsilon_l}{\varkappa_l}\frac{1 - \cosh \varkappa_l l }{\sinh \varkappa_l l} - \frac{\epsilon}{\varkappa}  \Big]{\cal A}_s,\\
    & -i\omega \theta_s + \frac{e^*}{\chi} \rho_s = -\Gamma E_0 \Big( q^2 \theta_s + \frac{2\pi i}{\Phi_0}q {\cal A }_s\Big),\, &
    j_s  = \Lambda \Big( \frac{\Phi_0}{2\pi} i q \theta_s - {\cal \bm A}_s\Big) + \sigma_n^{\rm L} \Big( i\omega\mathcal{A}_s - \frac{i q \rho_s}{\chi}\Big).
\end{align*}
To get the spectrum of the symmetric mode, one can solve this system of equations numerically. We turn to analyze two limits of the most physical interest. First, in the limit $ql \to 0$, we obtain:
\begin{align*}
    \omega_s(q) \approx - \frac{i q \sigma_n^{\rm L}  }{2\epsilon \epsilon_0}  \pm \sqrt{\frac{q \Lambda}{\epsilon \epsilon_0} - \Big( \frac{q \sigma_n^{\rm L}}{2\epsilon \epsilon_0}  \Big)^2},
\end{align*}
where we have used additional approximations analogous to the monolayer case: $\chi^{-1} = 0$ and the $O(q^2)$ term in the dynamics of $\theta_s$ can be neglected at low momenta.
For such small momenta, one can neglect the separation between the layers, giving effectively twice stronger Coulomb forces and resulting in frequency being $\sqrt{2}$ times larger than in the monolayer case. In the limit $q \to 0$, $l \to \infty$, and $\epsilon_l = \epsilon$, we reproduce the monolayer result:
\begin{align*}
    \omega_s(q) \approx \pm \sqrt{\frac{q  \Lambda}{2\epsilon \epsilon_0} - \Big( \frac{q \sigma_n^{\rm L}}{4\epsilon \epsilon_0}  \Big)^2} - i\left(  \frac{ q \sigma_n^{\rm L}}{4\epsilon \epsilon_0} \right).
\end{align*}

\subsubsection{Antisymmetric mode}

We now turn to investigate the antisymmetric combinations: $\theta_a = \theta_1 - \theta_2$, $\rho_a = \rho_1 - \rho_2$, $j_a = j_{1,\parallel} - j_{2,\parallel}$, ${\cal A}_a = {\cal A}_{1,\parallel} - {\cal A}_{1,\parallel}$. We derive the following closed set of equations:
\begin{align*}
    & \rho_a  = - \epsilon_0 q \omega \Big[ \frac{\epsilon_l}{\varkappa_l}\frac{1 + \cosh \varkappa_l l }{\sinh \varkappa_l l} + \frac{\epsilon}{\varkappa} \Big] {\cal A}_a  - \frac{2 i\omega \epsilon_l }{c^2\varkappa_l^2}J_z, & j_a  = - \epsilon_0 \omega^2 \Big[ \frac{\epsilon_l}{\varkappa_l}\frac{1 + \cosh \varkappa_l l }{\sinh \varkappa_l l} + \frac{\epsilon}{\varkappa} \Big]{\cal A}_a  - \frac{2i q}{\varkappa_l^2}J_z,\\
    & \vartheta = \frac{2 \pi \mu_0 l }{ \Phi_0 \varkappa_l^2}J_z - \frac{2\pi iq}{\Phi_0 \varkappa_l^2} {\cal A}_a, &
    J_z = J_0  (\theta_a - \vartheta) + \frac{i\omega \Phi_0}{2\pi \rho_c l}\vartheta, \\
    &    -i\omega \theta_a + \frac{e^*}{\chi} \rho_a = -\left(\frac{\Gamma E_0}{\lambda_J^2}\right) \Big[ \lambda_J^2 \Big( q^2 \theta_a + \frac{2\pi i}{\Phi_0}q {\cal A }_a\Big)  + 2(\theta_a - \vartheta) \Big], &
    j_a  = \Lambda \Big( \frac{\Phi_0}{2\pi} i q \theta_a - {\cal \bm A}_a\Big) + \sigma_n^{\rm L}\Big( i\omega\mathcal{A}_a - \frac{i q \rho_a}{\chi}\Big).
\end{align*}
The spectrum of the antisymmetric modes can be obtained by numerically solving this system of equations. Let us now analyse the limit $q\to 0$ to the leading order in $J_z$, setting $\rho_c^{-1} = 0$:
\begin{align}
    \omega_a(q = 0) \approx \sqrt{  \frac{2 e^* J_0}{\chi} + \frac{\omega_{ab}^2}{\epsilon_l \gamma^2} - \left(\frac{\Gamma E_0}{\lambda_J^2}\right)^2 }- i\frac{\Gamma E_0}{\lambda_J^2} ,
\end{align}
i.e. the antisymmetric modes are coherent gapped excitations, with the gap size being much larger than the lifetime. 

\subsection{Calculation of reflection coefficient}
In this subsection, we outline how to calculate the reflection coefficient $r_p(\q,q_z,\omega)$ for p-polarized waves incident on a Josephson-coupled superconducting bilayer. 
The steps are nearly identical to the derivation of longitudinal collective modes presented earlier in this appendix, and hence we just provide the essential steps.
We start with the vector-potential corresponding to the incident, reflected, and transmitted waves:
\begin{align}
 A_\parallel(\bm q, z;\omega) &  =  A_0
\begin{cases}
    e^{-iq_z^\epsilon (z-l/2)} - r_{p}  e^{iq_z^\epsilon (z-l/2)}, & z > \frac{l}{2}\\
 t_{p} e^{-iq_z^\epsilon (z+l/2)}, &  z < -\frac{l}{2}
\end{cases}\\
A_z(\bm q, z;\omega) & = \frac{q A_0}{q_z^\epsilon} 
\begin{cases}
    e^{-iq_z^\epsilon (z-l/2)} + r_{p}  e^{iq_z^\epsilon (z-l/2)}, & z > \frac{l}{2}\\
 t_{p} e^{-iq_z^\epsilon (z+l/2)}, &  z < -\frac{l}{2}
\end{cases}
\end{align}
Here $A_0$ is the amplitude of the incoming wave, $q_z^\epsilon = \sqrt{\epsilon_l \omega^2/c^2 - q^2}$, and $t_p$ is the transmission coefficient. 
The generic solution to the Maxwell equations in the region $-\frac{l}{2} < z < \frac{l}{2}$ reads as (with $q_l = \sqrt{\epsilon_l \omega^2/c^2 - q^2}$)
\begin{align}
    & A_\parallel (\bm q, z;\omega) = \alpha e^{- i q_l z} + \beta e^{i q_l z},\\
    & A_z (\bm q, z;\omega)  = \frac{ q \alpha}{q_l}   e^{-iq_l z} - \frac{q \beta}{q_l} e^{iq_l z} - \frac{4\pi}{c q_l^2} J_z,
\end{align}
where the coefficients $\alpha(\bm q,\omega)$ and $\beta(\bm q,\omega)$ are obtained by matching the tangential components of the vector potential at the two boundaries:
\begin{align}
        & \alpha = \frac{ {\cal A}_{2,\parallel } e^{i q_l l/2} - {\cal A}_{1,\parallel} e^{- i q_l l/2} }{e^{i q_l l} - e^{-i q_l l} }, & \beta = \frac{ {\cal A}_{1,\parallel} e^{i q_l l/2} - {\cal A}_{2,\parallel } e^{-i q_l l/2} }{e^{i q_l l} - e^{-i q_l l} },
\end{align}
where ${\cal A}_{1,\parallel} = A_0(1 - r_p)$ and ${\cal A}_{2,\parallel} = A_0 t_p$. From this, we obtain (after linearizing the Josephson current):
\begin{align}
    \vartheta & = - \frac{2\pi \mu_0 l }{\Phi_0 q_l^2}J_z  -  A_0 \frac{2\pi iq}{\Phi_0q_l^2} (t_p + r_p - 1),\\
    J_z & = J_0 (\theta_1 - \theta_2 - \vartheta) + \frac{i\omega \Phi_0}{2\pi \rho_c l}\vartheta.
\end{align}
The remaining boundary conditions give:
\begin{align}
    \rho_1  =  i\omega q \epsilon_0 A_0 \Big[ \frac{i \epsilon_l}{q_l} & \frac{t_p - (1 - r_p)\cos q_ll}{\sin q_ll} 
    + \frac{\epsilon}{q_z^\epsilon} (1 + r_p) \Big] 
     +\frac{i\omega \epsilon_l}{c^2 q_l^2} J_z,\\
    \rho_2  = i\omega q \epsilon_0 A_0 \Big[ \frac{i \epsilon_l}{q_l} & \frac{1-r_p - t_p\cos q_ll}{\sin q_ll}
    - \frac{\epsilon}{q_z^\epsilon} t_p \Big] -\frac{i\omega \epsilon_l}{c^2 q_l^2} J_z.
\end{align}
Finally, the continuity equations~\eqref{eqn:cnt_1} and~\eqref{eqn:cnt_2} read as
\begin{align}
    j_{1,\parallel} = \frac{\omega \rho_1}{q} - \frac{i J_z}{q}, & & j_{2,\parallel} = \frac{\omega \rho_2}{q} + \frac{i J_z}{q}.
\end{align}
Using the above equations together with the time-dependent Ginzburg-Landau dynamics of the order parameter phases, cf. Eq.~\eqref{eq:tdlg_bl}, and the expressions~\eqref{eqn:in_plane_j} for the in-plane current densities, we compute the reflection coefficient $r_p (q, q_z, \omega)$ numerically.

\end{document}